\documentclass[fleqn,usenatbib]{mnras}

\usepackage{newtxtext,newtxmath}

\usepackage[T1]{fontenc}

\DeclareRobustCommand{\VAN}[3]{#2}
\let\VANthebibliography\thebibliography
\def\thebibliography{\DeclareRobustCommand{\VAN}[3]{##3}\VANthebibliography}

\usepackage{graphicx}	
\usepackage{amsmath}	

\usepackage{xcolor}

\newcommand{\risco}{$r_{\rm{isco}}$}

\newcommand{\nicer}{\textit{NICER}}

\newcommand{\swift}{{\it Swift}}

\newcommand{\ixpe}{\emph{IXPE}}

\newcommand{\rev}{}

\title[Untangling the Variability in Fairall\,9]{Untangling the Complex Nature of AGN Variability in Fairall\,9}

\author[S. Hagen et al.]{Scott Hagen,$^{1}$\thanks{E-mail: scott.hagen@durham.ac.uk}
Chris Done,$^{1}$
Edward M. Cackett,$^{2}$
Ethan R. Partington,$^{3}$
Rick Edelson,$^{4}$
\newauthor
Collin Lewin,$^{5}$
Erin Kara,$^{5}$
and Jonathan Gelbord$^{6}$
\\
$^{1}$Centre for Extragalactic Astronomy, Department of Physics, Durham University, South Road, Durham, DH1 3LE, UK\\
$^{2}$Department of Physics and Astronomy, Wayne State University, 666 W. Hancock St, Detroit, MI, 48201, USA \\
$^{3}$Institute of Astrophysics, FORTH, GR-71110 Heraklion, Greece\\
$^{4}$Eureka Scientific Inc., 2452 Delmer St. Suite 100, Oakland, CA 94602, USA \\
$^{5}$MIT Kavli Institute for Astrophysics and Space Research, MIT, 77 Massachusetts Avenue, Cambridge, MA 02139, USA \\
$^{6}$Spectral Sciences Inc., 30 Fourth Ave. Suite 2, Burlington, MA 01803, USA
}

\date{Accepted XXX. Received YYY; in original form ZZZ}

\pubyear{\the\year{}}

\begin{document}
\label{firstpage}
\pagerange{\pageref{firstpage}--\pageref{lastpage}}
\maketitle

\defcitealias{Hagen24a}{HDE24}

\begin{abstract}

The accretion flow in AGN is not well understood, motivating intensive monitoring campaigns of multiwavelength variability to probe its structure. One of the best of these is the 3 year optical/UV/X-ray approximately daily monitoring campaign on Fairall\,9, a fairly typical moderate accretion rate AGN. The UV lightcurve shows a clear increase over $\sim 50$ days between years 1 and 2, strongly coherent with the X-ray lightcurve rise.  This changes the average spectral energy distribution such that the disc component is stronger while the X-ray spectrum steepens, so that the total X-ray power remains roughly constant. Outside of this global change, we apply a Fourier resolved analysis to test stochastic models where intrinsic fluctuations in the UV disc propagate down into the hard X-ray emission region via both changing the seed photon flux for Compton scattering (short light travel timescale) and changing the electron density (longer propagation timescale). Unlike these models, the hard X-rays are not particularly well correlated with the UV, and also have the wrong sign in that the hard X-rays \rev{marginally} lead the UV fluctuations.  We show that this is instead consistent with uncorrelated stochastic fluctuations in both the UV (slow) and X-ray (fast), which are linked together only weakly via light travel time.
These variability properties, as well as the changes in the SED, has implications for our understanding of AGN structure and physics, as well as future monitoring campaigns. 

\end{abstract}

\begin{keywords}
accretion, accretion discs -- black hole physics -- galaxies: active -- galaxies: individual: Fairall\,9
\end{keywords}



\section{Introduction}

Active galactic nuclei (AGN) are powered by the accretion of material onto a supermassive black hole (SMBH). This is generally understood in the framework of the standard \citet{Shakura73} disc model, where the accreting material forms a relatively cool optically thick, geometrically thin, disc if it is sufficiently dense to thermalise. Here the energy dissipated within the disc gives a radial temperature profile, increasing towards the central black hole. The predicted spectral energy distribution (SED) is then a sum of black-body components, typically peaking in the UV/EUV for bright AGN. 

However, this standard model generally fails in reproducing the observed phenomenology. Specifically, the \citet{Shakura73} disc prescription predicts that changes in the mass-accretion rate through the disc should only occur on time-scales of several thousand years for SMBHs. Monitoring campaigns show significant stochastic variability on time-scales of months to years \citep[e.g][]{MacLeod10}. More dramatically, a subset of AGN generally referred to as changing-look, display extreme variability, likely corresponding to significant structural changes within the flow, on time-scales of a few years \citep{Noda18, Ruan19, Krishnan24}. This is in contradiction to standard disc theory \citep{Lawrence18}.

However, AGN SEDs are clearly more complex than a simple multi-colour black-body, as expected from a standard disc. Firstly, they always show X-ray emission as a high-energy X-ray tail \citep{Elvis94, Lusso16}, which cannot originate from a standard, thermal disc, as the temperatures are too cool. Secondly, the standard disc models generally do not match in detail to the \rev{shape of the} observed bright blue/UV continua observed for AGN \citep{Antonucci89, Lawrence18}, showing instead a near ubiquitous down-turn in the UV occurring before the predicted disc peak \citep{Laor14, Cai23}. This down-turn appears to link to an upturn in the X-ray spectrum below $\sim 1$\,keV, lying above the X-ray power-law continuum, commonly referred to as the soft excess \citep{Laor97,Porquet04}.

The \rev{high energy} X-ray \rev{continuum} displays strong variability on short time-scales, indicating that \rev{it} likely originates from a compact coronal structure \citep{Lawrence87, Ponti12b}, which Compton up-scatters seed photons from the disc to higher energies \citep{Sunyaev80, Haardt93, Haardt94}. However, the geometry of this corona is not well understood, with popular ideas including placing it as a radially extended, geometrically thick, structure within the flow \citep[e.g][]{Narayan95, Rozanska00a, Done07, Kubota18} and as vertically extended or compact structure at the base of the jet \citep[e.g][]{Henri97, Markoff05, Kara19}, though we note that recent X-ray polarimetry results from \ixpe\ favour a radially extended geometry \citep{Krawczynski22, Gianolli23, Ingram23}. Regardless of the precise geometry, the highly variable nature of the X-ray provides a potential solution for the observed stochastic optical/UV disc variability, as a fraction of these photons will illuminate and heat the disc giving a variable reprocessed component to the optical/UV SED \citep{Clavel92}. 

The soft X-ray excess is even less well understood than the hard X-ray corona. 
One potential origin is the relativistic blurring of atomic line features induced by the reflection of X-rays from the primary hot coronal continuum \citep{Crummy06}. However, this often has issues with the available energy-budget, with recent work suggesting that it cannot account for all of the soft-excess emission \citep{Porquet24a, Porquet24b}. Instead, an alternative origin is from a warm Corona, where the disc emission fails to thermalise, forming instead a warm optically thick plasma overlaying a passive disc that Compton scatters disc photons into the EUV and soft X-ray bandpass \citep[e.g][]{Mehdipour11, Done12, Mehdipour15, Kubota18, Petrucci13, Petrucci18}. This also provides a potential explanation for the variability, as the disc structure now differs significantly from standard theory \citep{Rozanska15, Jiang20, Kawanaka24}, giving the possibility of variability on significantly shorter time-scales. In reality, the soft excess is likely a combination of emission from a warm corona and X-ray reflection off the optically thick disc structure \citep{Xiang22, Ballantyne24}. 

The recent intensive monitoring campaigns provide the ideal test set to uncover the physics and structure of AGN accretion flows \citep{Edelson15, Edelson17, Edelson19, Edelson24, McHardy14, McHardy18, Fausnaugh16, Chelouche19, Cackett18, Hernandez20, Kara21, Vincentelli21}. These were originally motivated by the models where X-ray illumination of the disc was responsible for the optical/UV variability via reverberation \citep{Blandford82, Welsh91}. The resulting light travel time lags are shortest for the UV emitting 
inner disc, and longer for the optical emitting outer disc, predicting a wavelength dependent time-lag  $\tau \propto \lambda^{4/3}$ \citep{Collier99, Cackett07}, which 
can map the size of the accretion disc.

However, while the data do show wavelength dependent lags, almost all objects mapped so far show that the 
variability is significantly more complex than predicted from simple reverberation models. The lags are longer than expected for standard discs, but an even more serious problem is that the optical and UV variability is only poorly correlated with the X-rays which are meant to be driving it \citep[see e.g the compilation in][]{Edelson19}. The long lags can be explained if 
the reprocessing is mainly from 
the broad line region, or a wind on the inner edge of the broad line region
\cite[e.g][]{Korista01, Korista19, Dehghanian19b, Chelouche19, Kara21, Netzer22}. Alternatively the long lag time could indicate that the X-ray source is at larger distances from the disc, perhaps on the spin axis, 
\citep{Kammoun21b, Kammoun21a}. However, the poor correlation with the X-rays is harder to explain in all these scenarios. Work directly modelling the optical/UV light-curves through standard disc reverberation give light-curves that are extremely well correlated with the X-ray \citep{Gardner17, Mahmoud20, Mahmoud23, Hagen23a}, as the light-travel time to the inner disc is not long enough to smooth out the fast sub-day X-ray variability in order to give the slow optical/UV. 

These issues motivate alternative models for AGN variability. \citet{Panagiotou22b} show that the poor optical/UV to X-ray correlation can be re-produced if the X-ray source illuminating the disc is dynamically varying its height along the jet axis \citep[see also e.g][]{Panagiotou22a, Kammoun24}.
However, this does not explain recent results suggesting slow fluctuations moving through the disc itself \citep{Yao23, Neustadt22, Neustadt24}. This instead is more consistent with 
variability intrinsic to the disc itself, \rev{as suggested in earlier monitoring campaigns \citep{Uttley03, Arevalo08, Arevalo09}, and which} could be achieved if the structure deviates from standard disc theory, and appears to be required by recent radiation-MHD simulations \citep{Secunda24} showing an almost negligible impact to the UV variability through reprocessing. Recent models for this, using either temperature fluctuations within the disc \citep{Cai18, Cai20, Su24} or propagating mass-accretion rate fluctuations \citep{Lyubarskii97, Arevalo06, Hagen24a}, are able to generate light-curves that qualitatively match the data.

Fairall\,9 is a local clean Seyfert\,1 AGN, displaying little to no complex absorption signatures, and thus a direct view of the central engine \citep{Lohfink16}. It has recently been jointly monitored for $\sim 1000$\,days by \swift\ \citep{Hernandez20, Edelson24}, \nicer\ \citep{Partington24}, and the Las Cumbres Observatory (LCO) \citep{Hernandez20}. This makes it an ideal testbed for uncovering the nature of AGN variability. 
The culmination of these data show a serious departure from the standard reverberation picture. In the first year \citet{Hernandez20} showed a clear connection between the UV and X-ray, with lags increasing with the expected $\lambda^{4/3}$. However, their normalisation was significantly longer than \rev{predicted for disc reverberation}, instead being more consistent with a wind \citepalias{Hagen24a}. There were additional detections of inwards propagating lags on long time-scales \citep{Hernandez20, Yao23}, which is clearly in contradiction to the standard picture. While the first year data initially appeared to give a clear picture of the system, further monitoring in \citet{Edelson24} introduced additional complexity. Here the UV-X-ray lag appeared to vary throughout the campaign, occasionally switching sign. The analysis in \citet{Partington24}, while clearly displaying a soft excess component with near identical variability properties to the UV, also show tentatively the soft excess leading \emph{both} the X-ray coronal and UV disc variability (though we note their posterior on the X-ray lag is quite broad and the correlation is rather poor at $\sim 0.5$). Overall, this source displays a rather rich, and at times entirely confusing, phenomenology significantly more complex than the standard picture.

In this paper we aim to unravel the components responsible for driving the AGN variability, with a focus on the connection between the UV emitting disc and X-ray corona. Working from the predictions presented in \citet{Hagen24a} (hereafter \citetalias{Hagen24a}), the slow fluctuations within the disc propagating into the corona will lead to the UV leading the X-ray emission on long time-scales. Then the X-rays, which vary on much shorter time-scales, illuminate the disc, imprinting a reverberation signal where the X-ray leads the UV on fast time-scales. Hence the propagating mass-accretion rate fluctuations scenario comes with the key prediction of a switch from the UV leading to following the X-ray as a function of temporal frequency. This is seen in Fourier resolved lag measurements between X-ray light-curves in AGN \citep{Kara13a, Kara13b, Kara16}, and is generally attributed to fluctuations within the corona itself. However, as of yet there is no confirmed detection of \rev{the disc substantially leading the corona, consistent with predictions from propagating disc-corona fluctuations}, though we note it was suggested in \citet{Breedt09} as a possibility for the long term variability in Mrk\,79.

Here we focus on \rev{the} \swift\ \rev{monitoring data}, with an average cadence of $1-2$\,days, giving a range of near three orders of magnitude in temporal frequency. We will start by assessing the SED throughout the campaign (section\,\ref{sec:data}), as this gives an overview of the energy generating structure and its (possible) evolution throughout the campaign. We will then continue onto extracting the power-spectra as well as the Fourier resolved time-lags in section\,\ref{sec:fourier}, using Gaussian Process regression to obtain evenly sampled light-curves \citep[following the work of][]{Wilkins19, Griffiths21, Lewin22, Lewin23, Lewin24}. In this section we also present an analytic approach to modelling the Fourier resolved timing signatures, used to give a framework for interpreting the system timing signatures in the context of an intrinsically variable disc and corona. In light of both the SEDs and the timing-signatures, we speculate that the system is likely \rev{undergoing} an evolution in the disc component driven by an increase in the global mass-accretion rate. It is also clear that current models for AGN variability do not adequately explain the observed phenomenology. We give these conclusions in section\,\ref{sec:conc}.

To avoid confusion between spectral and temporal frequency, we stress that in this paper we will only be referring to frequency as \emph{temporal frequency}, $f$. Any reference to energy spectra or SEDs will be given in terms of energy, $E$.

\section{Light-curves and SED analysis}
\label{sec:data}

\begin{figure*}
    \centering
    \includegraphics[width=\textwidth]{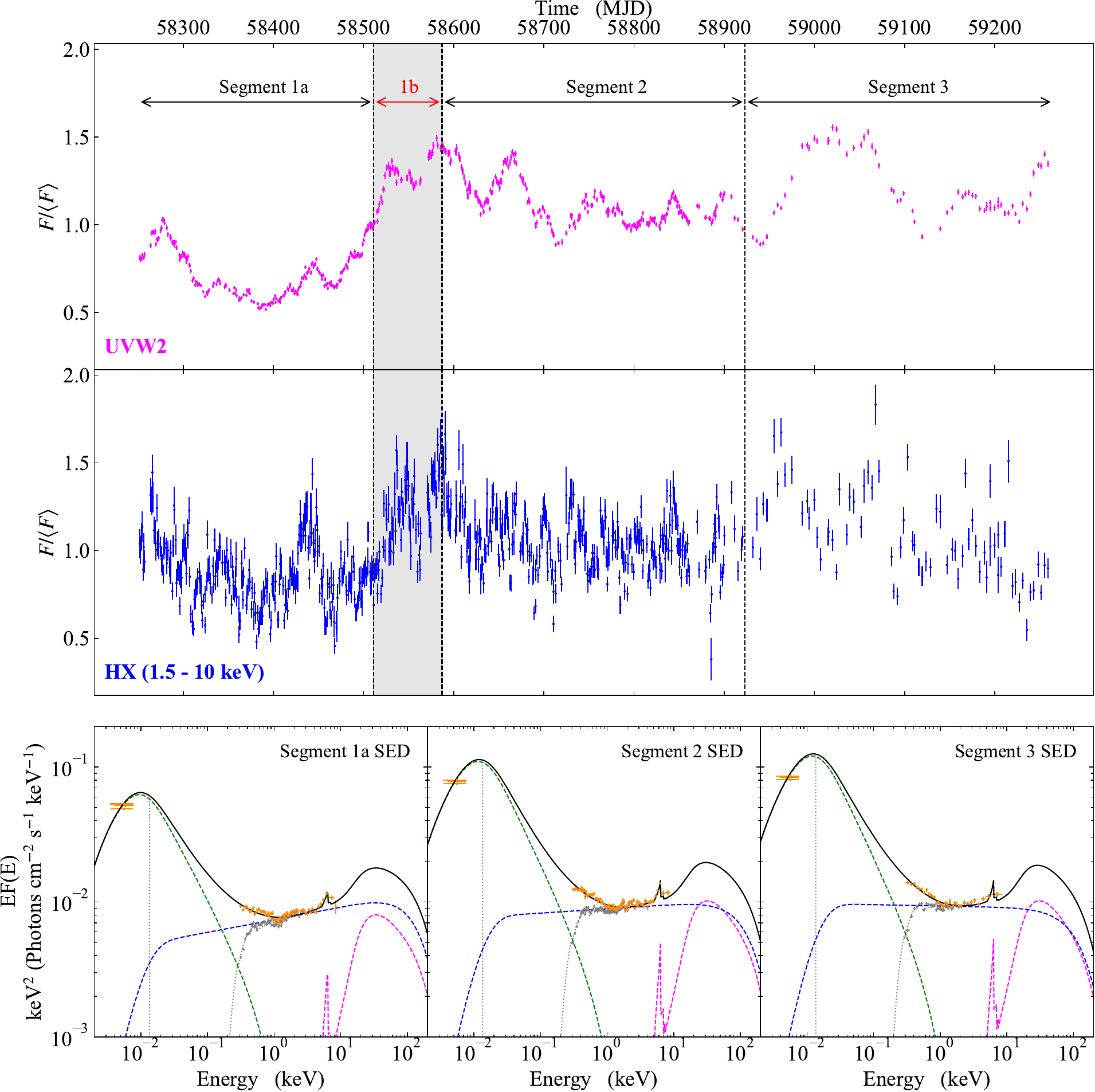}
    \caption{
    \textbf{Top rows:} Light-curves for Fairall\,9 for \swift\ UVW2 (top, magenta) and HX (1.5-10.0\,keV, bottom, blue) \rev{taken from \citet{Edelson24}. These are shown as fractional flux to allow for a more direct comparison, given that the native units of each light-curve differ (cts/s for HX and erg s$^{-1}$ cm$^{-2}$ \AA$^{-1}$ for UVW2)}. The dashed lines indicate the segments used to create time-averaged SEDs. The shaded grey region indicates a sharp rise in the light-curves, which we have chosen to exclude from the SED analysis due to the possibility of \rev{a change in the inner geometry}. \\
    \textbf{Bottom row:} SEDs averaged over each of the segments indicated in the light-curves. The orange data-points show the unfolded, de-absorbed, data, while the grey points indicate the absorbed data. \rev{We note that the x-errorbars in the UVOT data represent the full band-pass. During the fitting procedure the filter response is also taken into account.} The solid black line shows the total SED model, with the dashed lines indicating the model components to the total SED. These are: a warm Comptonising disc (green), a hot Comptonising corona (blue), and neutral reflection of optically thick material (magenta).
    }
    \label{fig:LC_and_SEDs}
\end{figure*}

\subsection{Extracting the SEDs}

We begin our analysis by examining the long-term evolution in the spectral energy distribution (SED). Here we are mainly interested in large changes to the SED. Therefore we divide the light-curves into three segments, shown as the dashed vertical lines in Fig.\ref{fig:LC_and_SEDs}, and create time-averaged SEDs within each segment. This will average out rapid stochastic variations on time-scales much shorter than the segment length, effectively removing the fast intrinsic X-ray variations and the expected reverberation signal in the optical/UV. 

We initially divide the light-curves into roughly equal segments of 336\,days. This gives a natural division between the lower flux state from the first year (segment\,1) of the campaign compared to the higher flux states later on (segments\,2 and 3), as well as a division between the high cadence (segments\,1 and 2) and low cadence (segment\,3) part of the campaign. However, visual examination of the light-curves show a clear sharp rise in the time-interval from 58511\,MJD to 58590\,MJD in both the UV and X-ray light-curves, indicated by the grey shaded region in Fig.\ref{fig:LC_and_SEDs}. Due to the large difference in average UV flux between segment\,1 and segments\,2 and 3, this rise could potentially indicate a more significant change to the accreting system, which would then significantly alter the shape of the time-averaged SED. Hence, we have elected to exclude this region from the SED analysis, defining segment\,1a to end at 58511\,MJD. If there are no significant changes in SED shape during the rise, then this choice will have no meaningful effect, whereas if there is a change then this should be visible in a comparison between the segmented SEDs.

The UV portion of the SED is constructed from the \swift-UVOT filters UVW1, UVM2, and UVW2. Lower energy filters are ignored to avoid significant contamination from the host galaxy. We start by calculating the average flux in each bandpass for each segment, using the light-curves presented in \citet{Edelson24} (we also refer the reader to this paper for details regarding the data reduction). These fluxes are then de-reddened using the extinction law from \citet{Fitzpatrick07} and an $E(B-V) = 0.0223$ taken from the dust maps of \citet{Schlafly11}\footnote{Extracted from: \url{https://irsa.ipac.caltech.edu/applications/DUST/}}. Finally, these are then converted to a count-rate, using the conversion factors in \citet{Poole08}, and written to {\sc xspec} readable files along with their corresponding filter-response curves.

The X-ray component of the SED consists of time-averaged \swift-XRT data between 0.3\,keV and 10.0\,keV. We use the \swift-XRT analysis tools described in \citet{Evans09}, integrated in the standard {\sc swiftools} API\footnote{\url{https://www.swift.ac.uk/API/}}, to download, extract and stack spectra within each segment. As outlined in \citet{Evans09} this will, for each observation, check for pile-up, identify the source, automatically select an ideal extraction annulus based on the source count-rate, and then generate an event list, source spectrum, and ancillary response file (ARF). Each observation is then combined into a single time-averaged spectrum for the segment. While adding together the ARFs, a weighting is applied based on the proportion it contributes to the total counts. As well as an averaged source spectrum, the time-averaged background spectrum is generated using an annulus with inner width $142''$ and outer width $260''$ centred on the source for each observation. For a full description of the data reduction pipeline we refer the reader to \citet{Evans09}. Finally, we re-bin the time-averaged spectrum for each segment such that each bin contains at least 20\,counts, allowing for the use of $\chi^{2}$ statistics.

\subsection{Modelling the SEDs - Theoretical Overview}


When modelling the SED, and throughout the rest of this paper, we will use the standard notation for radii, such that $R$ is radius in physical units while $r$ is in dimensionless gravitational units, with $R = r R_{G}$ and $R_{G} = GM/c^{2}$. Similarly, for the mass-accretion rate $\dot{M}$ indicates physical units, while $\dot{m}$ is the dimensionless mass-accretion rate scaled by the Eddington mass-accretion rate, such that $\dot{M} = \dot{m} \dot{M}_{\rm{Edd}}$. Here we have defined $\dot{M}_{\rm{Edd}}$ such that $L_{\rm{Edd}} = \eta(a_{*}) \dot{M}_{\rm{Edd}} c^{2}$, where $\eta(a_{*})=\sqrt{1 - 2/(3r_{\rm{isco}})}$ is the spin-dependent radiative efficiency.

To identify physically meaningful changes to the SED between segments we apply the {\sc agnsed} model from \citet{Kubota18} to the time-averaged data from each segment. Briefly, {\sc agnsed} assumes a radially stratified flow, subdivided into an outer standard \citet{Shakura73} disc, an intermediary warm Comptonisation region (still assuming a disc geometry), and an inner geometrically thick hot Comptonisation region. The model assumes that each of these regions has a standard radial emissivity profile, $\epsilon(R) \propto R^{-3} f(R)$, where $f(R)$ describes the stress free inner boundary for a Kerr black hole \citep{Novikov73, Page74}. As in \citet{Hagen23a} and \citetalias{Hagen24a} we will assume the entire disc is covered by the warm Comptonising corona, effectively reducing the flow to two zones only: the warm Comptonisation region and the hot Comptonisation region. We give here a brief description of each of these, while referring the reader to \citet{Kubota18} for a detailed description. 

\textit{Warm Comptonisation region:} This region considers the case where the disc fails to thermalise. Instead, the dissipation region moves away from the mid-plane into the photosphere, forming a slab geometry above a passive dense disc structure \citep{Petrucci13, Petrucci18}. The underlying disc re-processes the power dissipated in the photosphere, which then sets the seed photons that are Comptonised. This then self-consistently sets the seed-photon temperature at any given radius to the expected disc temperature $T_{NT}(R)$, which results in the seed-photon temperature increasing with decreasing radius. The electron temperature, $kT_{e, w}$, which sets the high energy roll over, is assumed constant across the entire region, and is left as a free parameter throughout. The spectral slope is set by the photon index, $\Gamma_{w}$, which we also leave free throughout. An important side-effect of optically thick warm Comptonisation with a \emph{passive} underlying disc is that it gives a fairly narrow range in the expected photon index of $\Gamma_{w} = 2.5 - 2.7$ \citep{Petrucci18}. Significant softening beyond this would indicate an increase in seed-photon power, which would require some dissipation in the mid-plain (i.e a non-passive disc).

\textit{Hot Comptonisation region (corona):} Below a radius $r_{h}$ the warm Comptonising region evaporates into a hot geometrically thick flow, responsible for the high energy X-ray tail \citep{Narayan95, Liu99, Haardt93, Haardt94}. This flow is assumed to extend down to \risco. The power dissipated in the flow between $r_{h}$ and \risco, $L_{h, \rm{diss}}$, sets the power intrinsic to the corona. However it also sees seed photons intercepted from the disc, $L_{h, \rm{seed}}$, such that the total power of the hot Comptonisation region is $L_{h} = L_{h, \rm{diss}} + L_{h, \rm{seed}}$. As with the warm Comptonising region the spectral shape is set by the seed photon temperature, $kT_{\rm{seed}, h}$, the electron temperature, $kT_{e, h}$, and the photon index $\Gamma_{h}$. The seed photon temperature is self-consistently calculated internally, \rev{assuming} these originate from the warm Comptonisation region. The electron temperature we fix at $kT_{e, h} = 100$\,keV as we lack sufficient high-energy coverage to constrain this. The photon index, $\Gamma_{h}$, we leave free. 
Theoretically, this should be set by the Compton balance \citep[e.g][]{Beloborodov99} (i.e the ratio of $L_{h, \rm{seed}}$ to $L_{h, \rm{diss}}$), \rev{and so we also expect to see it vary if there is variability in the power intrinsic to the disc/warm Comptonisation region relative to that of the hot Corona \citepalias[see e.g][]{Hagen24a}}

A key aspect of the {\sc agnsed} suite of models is the energy balance. The total power available to emit is fundamentally set by the black hole mass, $M$, the Eddington scaled mass-accretion rate, $\dot{m}$. The black hole spin, $a_{*}$, also contributes by setting the radiative efficiency and inner radius, \risco, of the flow. Thus, {\sc agnsed} gives an estimate of the relative amount of power dissipated within each region. In the model context this is set by adjusting the radial extent of each component, giving a (highly model dependent) size estimate for each emitting component. While the exact values this gives should not be treated as precise measurements, significant changes in the size scales between light-curve segments are indicative of more systematic changes to the accretion structure itself.

\subsection{Modelling the SEDs - Results}
\label{sec:SED_results}

Using {\sc xspec} v. 12.13.0c \citep{Arnaud96}, we now fit each segment. We fix the black hole mass to $M = 2 \times 10^{8}\,M_{\odot}$ and co-moving distance of $200$\,Mpc \citep{Bentz15}. As Fairall\,9 is a `clean' AGN, displaying little to no intrinsic absorption \citep{Lohfink16}, we assume the systems inclination must be relatively low, and set $\cos(i) = 0.9$. {\sc agnsed} does not include any relativistic ray-tracing, hence we fix the black hole spin to $0$, as in this regime the expected general-relativistic corrections to the SED are minimal \citep[see][]{Hagen23b}. Finally, we fix the outer disc radius to the self-gravity radius, $r_{sg}$, from \citet{Laor89}, and only consider the case where the entire disc is covered by the warm Comptonising medium. The coronal height, $h$, which is used as a tuning parameter for the re-processed signal, is fixed at $10\,R_{G}$.

In addition to {\sc agnsed} we include a {\sc pexmon} component \citep{Magdziarz95, Nandra07} to model the Fe-K$\alpha$ emission, convolved with {\sc rdblur} \citep{Fabian89} to account for any smearing in the reflection spectrum. We tie the spectral index of {\sc pexmon} to that of the hot Comptonised component in {\sc agnsed}, and fix all abundances to their solar values. The only free parameter here is the normalisation. All parameters of {\sc rdblur} are either fixed or tied to their {\sc agnsed} counterparts (see Table\,\ref{table:SEDpars}).

Finally, we include a global photoelectric absorption component, {\sc phabs}, to account for Galactic absorption. Here we fix the column-density to $N_{H} = 3.5 \times 10^{20}\,\rm{cm}^{-2}$ \citep{HI4PI16}. The total model, in {\sc xspec}, syntax is then: {\sc phabs*(agnsed + rdblur*pexmon)}, where we fix the abundances to those of \citet{Anders89} and the cross-sections to those of \citet{Verner96}.

\begin{figure}
    \centering
    \includegraphics[width=\columnwidth]{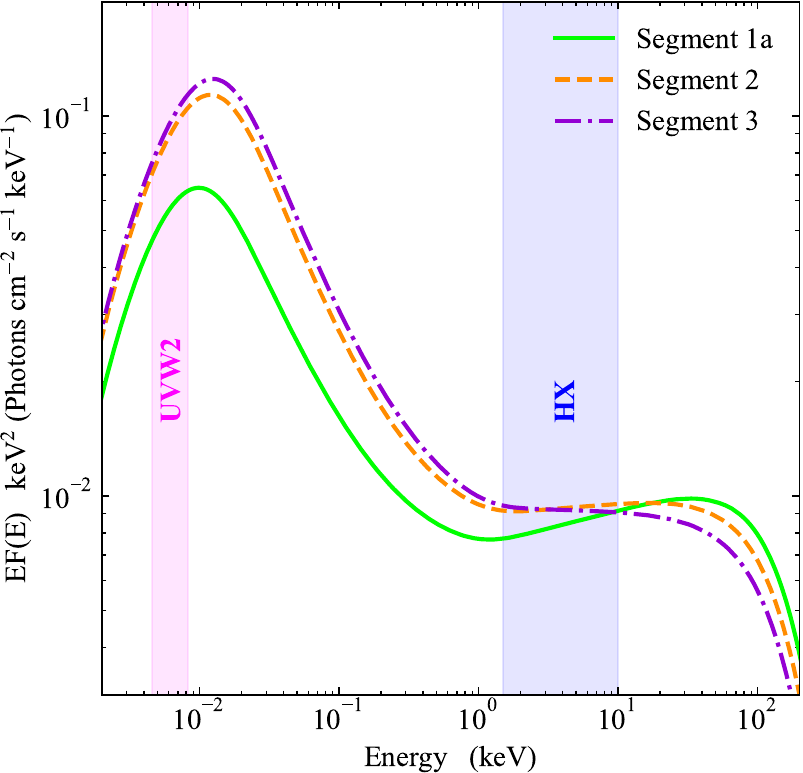}
    \caption{
    A comparison of the predicted intrinsic SEDs for each light-curve segment highlighted in Fig.\,\ref{fig:LC_and_SEDs}. Here the solid green line shows segment\,1a, the dashed orange line shows segment\,2, and the dashed-dotted purple line shows segment\,3. The shaded regions highlight the \swift\ UVW2 and HX band-passes. It is clear that the majority of the variability between segments occurs in the disc-like region. The time-averaged, integrated, hot Coronal component displays little variability, suggesting that the long term variability in the HX band-pass is mainly due to the change in photon-index. 
    }
    \label{fig:SEDcomp}
\end{figure}

The SED from each segment is fit separately, to allow for variability in the intrinsic SED parameters. We use the {\sc xspec} MCMC capability, and run each each fit for 300\,000 steps after an initial burn-in of 30\,000 steps using 6 walkers and a Goodman-Weare algorithm. The fit results are given in Table\,\ref{table:SEDpars}, and the corresponding model SEDs are shown in the bottom row of Fig.\ref{fig:LC_and_SEDs}. We also show the dsitributions for the {\sc agnsed} parameters in Fig.\,\ref{fig:SED_parDist} (for full contour plots see Appendix\,\ref{app:SED_contour}) and the resulting {\sc agnsed} \emph{only} models for each segment in Fig.\,\ref{fig:SEDcomp}.

There is a clear evolution in the SED between segment\,1a and segments\,2 \& 3. This is most easily seen in Fig.\,\ref{fig:SEDcomp}, where we show the best fit {\sc agnsed} model for each segment. The optical/UV emitting disc is markedly lower in segment\,1a, which also corresponds to a significantly lower mass-accretion rate (Fig.\ref{fig:SED_parDist}). An interesting point to note, however, is that although the optical/UV clearly changes, the integrated X-ray power is markedly close to constant. It appears that the majority of the HX variability between segments is mainly due to changes in the photon index, which softens as the mass-accretion rate increases.

\begin{figure*}
    \centering
    \includegraphics[width=\textwidth]{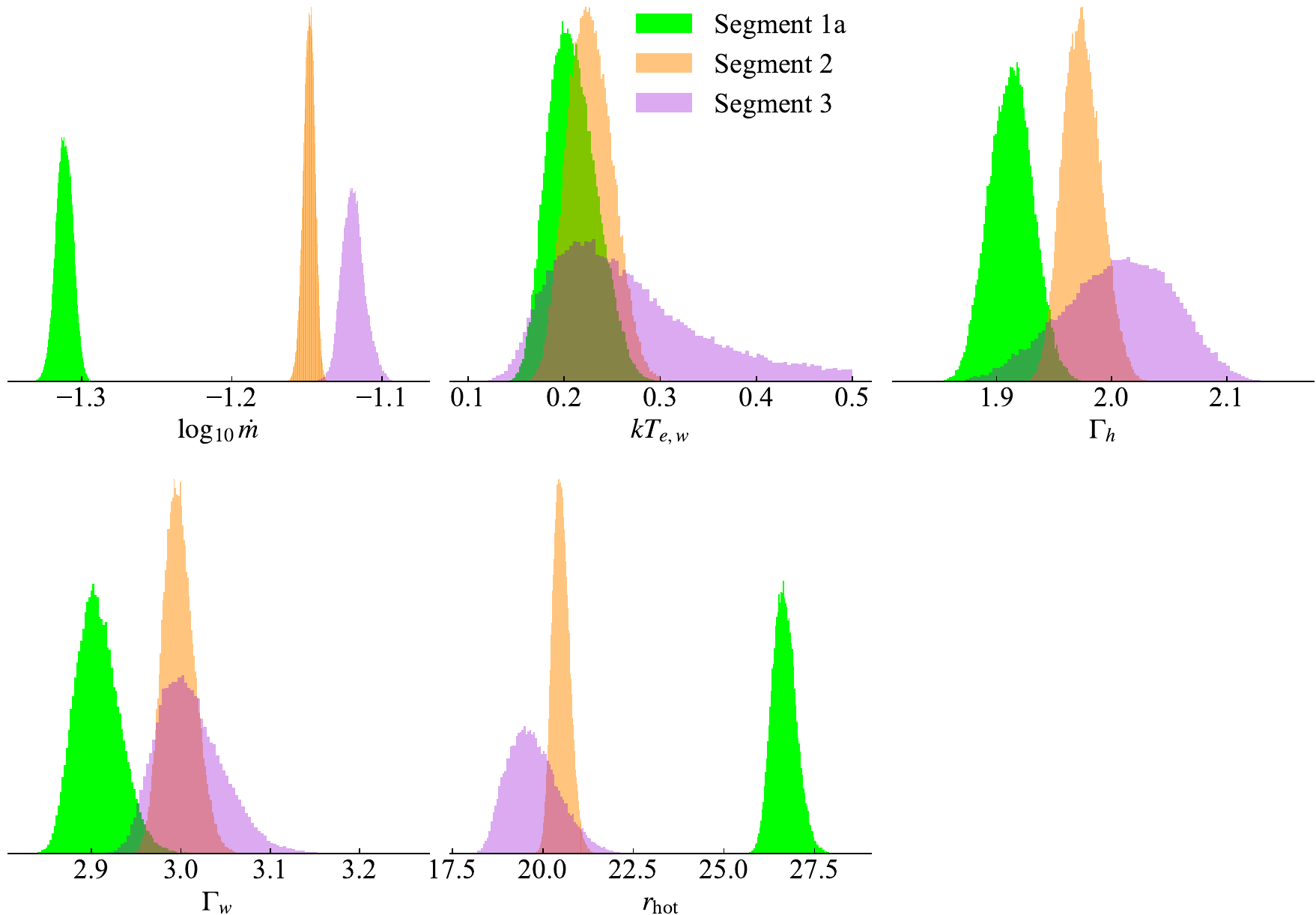}
    \caption{
    Posterior distributions for the free {\sc agnsed} parameters from our MCMC fit, normalised such that they integrate to 1. Here green corresponds to the segment 1a SED, orange to the segment 2 SED, and purple to the segment 3 SED. There is clearly a significant increase in mass-accretion rate between segment 1a and segments 2 and 3, corresponding to before and after the strong rise in the UVW2 light-curve. There is also a significant softening of the hot corona photon index, suggesting an increase in seed photon flux entering the corona leading to increased Compton cooling. Additionally we see a softening in the warm corona photon index, which suggests an increase in power dissipated within the mid-plain. The reduction in truncation radius, $r_{\rm{hot}}$, is a consequence of the energy balance compensating for the increased mass-accretion rate (seen in the optical/UV) while the integrated X-ray power remains remarkably constant. The segment\,3 posteriors are generally broader than in segments\,1a \& 2 as during this time the cadence of the campaign was reduced to $\sim 4$\,days, resulting in fewer observations in our stack and thus lower S/N.
    }
    \label{fig:SED_parDist}
\end{figure*}

The segmented SEDs average out the fast variability, and so these are only representative of changes on time-scales of hundreds of days. In that context, the seeming lack of change in the intrinsic X-ray power while the disc-like power increases is challenging to reconcile with both a pure reverberation picture and a propagation scenario. Significant changes in the optical/UV through reverberation naturally requires significant changes in the emitted X-ray power. Clearly this is not the case here. For propagation, if fluctuations travel from the disc to the corona, then one would expect to see an increase in power in both components. \rev{This does not appear to be the case here.} 

{\renewcommand{\arraystretch}{1.6} 
\begin{table*} 
    \centering 
    \begin{tabular}{c c c c c} 
   \textbf{Component} & \textbf{Parameter (Unit)} & \textbf{Segment 1a} & \textbf{Segment 2} & \textbf{Segment 3} \\ 
   \hline 
   \hline 
   {\sc phabs} & $N_{H}$ ($10^{22}\,\rm{cm}^{-2}$)  & 0.035  & 0.035  & 0.035 \\ 
   \hline 
   {\sc agnsed} & $M$ ($M_{\odot}$)  & $2 \times 10^8$  & $2 \times 10^8$  & $2 \times 10^8$ \\ 
   & Dist (Mpc)  & 200  & 200  & 200 \\ 
   & $\log_{10} \dot{m}$ ($\dot{M}/\dot{M}_{\rm{Edd}}$)  & $-1.310^{+0.008}_{-0.011}$  & $-1.148^{+0.005}_{-0.007}$  & $-1.12^{+0.01}_{-0.01}$ \\ 
   & $a_{*}$  & 0  & 0  & 0 \\ 
   & $\cos(i)$  & 0.9  & 0.9  & 0.9 \\ 
   & $kT_{e, h}$ (keV)  & 100  & 100  & 100 \\ 
   & $kT_{e, w}$ (keV)  & $0.20^{+0.05}_{-0.03}$  & $0.23^{+0.04}_{-0.04}$  & $0.22^{+0.21}_{-0.05}$ \\ 
   & $\Gamma_{h}$  & $1.91^{+0.04}_{-0.03}$  & $1.97^{+0.03}_{-0.02}$  & $2.01^{+0.07}_{-0.09}$ \\ 
   & $\Gamma_{w}$  & $2.90^{+0.05}_{-0.03}$  & $2.99^{+0.03}_{-0.02}$  & $3.00^{+0.08}_{-0.04}$ \\ 
   & $r_{h}$  & $26.6^{+0.6}_{-0.4}$  & $20.5^{+0.4}_{-0.3}$  & $19.7^{+1.3}_{-1.0}$ \\ 
   & $r_{w}$  & = $r_{\rm{out}}$  & = $r_{\rm{out}}$  & = $r_{\rm{out}}$ \\ 
   & $\log_{10} r_{\rm{out}}$  & = $r_{sg}$  & = $r_{sg}$  & = $r_{sg}$ \\ 
   & $h_{\rm{max}}$  & 10  & 10  & 10 \\ 
   \hline 
   {\sc rdblur} & Index  & -3  & -3  & -3 \\ 
   & $r_{\rm{in}}$  & = $r_{h}$  & = $r_{h}$  & = $r_{h}$ \\ 
   & $r_{\rm{out}}$  & = $r_{w}$  & = $r_{w}$  & = $r_{w}$ \\ 
   & Inc.(deg)  & 25.8  & 25.8  & 25.8 \\ 
   \hline 
   {\sc pexmon} & $\Gamma$  & = $\Gamma_{h}$  & = $\Gamma_{h}$  & = $\Gamma_{h}$ \\ 
   & $E_{c}$ (keV)  & 1000  & 1000  & 1000 \\ 
   & Inc. (deg)  & = Inc.  & = Inc.  & = Inc. \\ 
   & Norm ($10^{-3}$)  & $6.1^{+3.4}_{-1.7}$  & $9.8^{+3.2}_{-1.7}$  & $11.5^{+8.3}_{-6.5}$ \\ 
   \hline 
   \hline 
   $\chi^{2}$/d.o.f &  & 723.13/567 = 1.28  & 761.0/624 = 1.22  & 388.52/414 = 0.94 \\ 
   \hline 
\end{tabular} 

    \caption{
    Best fit parameters for the SED model for each segment. Column 1 (from the left) gives the model component. Column 2 gives the model parameter, along with the relevant units inside the brackets. If no units are given then the parameter is dimensionless. Columns 3, 4, and 5 give the best fit parameters for segments 1a, 2, and 3 respectively. The errors are at 90\,\% confidence, calculated from our MCMC chain. Parameters with no errors were frozen during the fit procedure. Parameter values prefixed by an `$=$' (e.g $= \Gamma_{h}$) indicate that they were tied to the parameter immediately following. The exception here is $\log_{10} r_{\rm{out}}$, which was set to the self-gravity radius throughout (calculated from \citealt{Laor89}). Not listed, we also fix the redshift to $z = 0.045$ and all abundances to the solar values.
    }
    \label{table:SEDpars} 
\end{table*}
}

Instead, it appears the only coherent signal between the optical/UV and HX on these time-scales occurs through the balance between Compton cooling and heating within the hot flow. The Compton heating of the electrons fundamentally depends on the power dissipated within the corona, $L_{\rm{diss}}$. 
\rev{The lower energy seed photons from the disc, with power $L_{\rm{seed}}$, then work against this heating by cooling the electrons through inverse Compton scattering}.
This balance between heating and cooling sets the expected photon index from Comptonisation \citep{Beloborodov99}. If there is increased power dissipated in the corona, while the seed photon power remains constant, \rev{the electrons will be more efficiently heated} predicting a harder spectrum. Conversely, if the seed-photon power increases while the dissipated power remains constant you will more efficiently cool the electrons, predicting a softer spectrum. This appears to be the case in Fairall\,9 between segments, with a significant softening of $\Gamma_{h}$ occurring in tandem with a significant brightening of the disc.

On the whole it appears that the large changes seen in the light-curves are primarily driven by changes to the disc-like region. In the context of {\sc agnsed} this coincides with a $\sim 40\,\% - 50\,\%$ increase in the global mass accretion rate. It also appears to require a softening in the photon index for the soft X-ray excess, now well beyond what is expected for a passive underlying disc \citep{Petrucci13}. Instead, this suggests that as the mass-accretion rate increases an increasing fraction of power is dissipated towards the mid-plane, increasing the effective seed-photon power and softening the warm corona.

As the integrated X-ray power remains roughly constant, the increase in global mass-accretion rate requires an increase in the fraction of available power dissipated in the disc-like region. {\sc agnsed} preforms this balance by reducing the radius of the hot corona, formally moving the disc further in. While this suggests \rev{an evolution in the structure of the inner flow}, this interpretation is \rev{naturally} model dependent. However, it does fit into the results of \citet{Edelson24}, who showed an evolution in both the lag and overall correlation between the X-ray and UV throughout the campaign, which is not expected for a \rev{static} system \rev{(i.e no evolution in its inner geometry)}.  Their results, however, were based on cross-correlation techniques, which do not account for signal potentially occurring on different time-scales. Hence, to uncover the potentially evolving nature of Fairall\,9 we revisit the timing-properties of the data, but using a Fourier resolved approach, in order to isolate separate physical processes within the data.

\section{Fourier Analysis}
\label{sec:fourier}

\subsection{Gaussian Process Regression}

\begin{figure*}
    \centering
    \includegraphics[width=\textwidth]{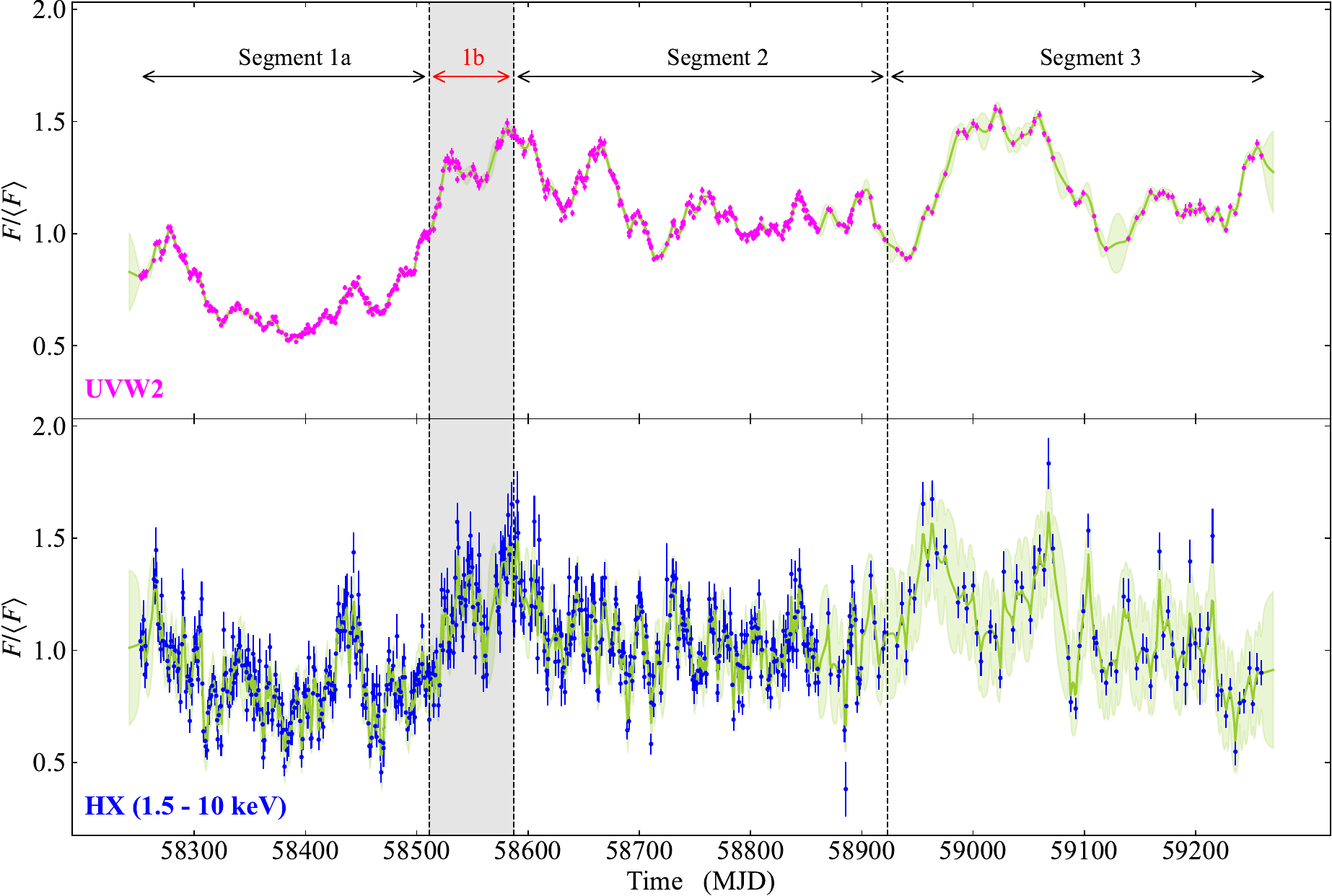}
    \caption{
    \swift\ UVW2 (\textbf{top}) and HX (\textbf{bottom}) light-curves, including our Gaussian Process (GP) model. The mean GP prediction is shown as the solid green line in both light-curves, while the $1\sigma$ dispersion is shown by the green shaded regions. As in Fig.\ref{fig:LC_and_SEDs} the dashed lines indicate the segments used for our SED and now Fourier analysis. Here we also include the segment 1b for the Fourier analysis, in order to evaluate the impact of the sharp rise to the variability. 
     }
    \label{fig:GP_LCs}
\end{figure*}

The estimation of Fourier based timing products (e.g power-spectra, time-lags, etc) revolve around the use of discrete fast Fourier transforms (hereafter FFT). This technique has the advantage that it by definition separates out variability on different time-scales, potentially isolating separate physical mechanisms if they are present. However, FFTs require evenly sampled data to work, which real data are not.  There are currently several methods to circumvent this issue, pioneered by previous works. The two most commonly used in AGN reverberation studies are the maximum likelihood inference technique \citep{Miller10, Zoghbi13a, Zoghbi13b} and Gaussian Process regression \citep{Wilkins19, Lewin22, Lewin23, Lewin24}. The maximum likelihood approach involves directly fitting the power spectral density (PSD) to the time-series, by transforming the the PSD to the time-domain, giving the auto correlation function, which is used to populate a covariance matrix for the data that then leads to a likelihood. The Gaussian process approach directly models the light-curves with a Gaussian Process regressor, which is then used to draw evenly sampled light-curve realisations, allowing for the direct use of FFTs. In this paper we opt for the Gaussian Process (hereafter GP) method. Here we give a brief overview of the GP model employed in our paper, and the resulting Fourier products. As the data are already very well sampled, we also present a re-production of the following results using simple linear interpolation as a sanity check in Appendix\,\ref{app:lin_interp}.

We refer the reader to \citet{Rasmussen06} for a comprehensive overview of GPs, and to \citet{Wilkins19, Griffiths21, Lewin22} for a comprehensive description of their application to AGN light-curves. Here we limit ourselves to a brief description relevant to this paper, rather than a full outline of the concepts behind GPs.

Throughout we use the framework implemented in the {\sc python} package {\sc GPy} v.1.10.0 \citep{GPy14}\footnote{\url{https://gpy.readthedocs.io/en/deploy/}}. In short GPs assume that our data have been drawn from a multivariate normal distribution, with mean function $m(t) = \mathbb{E}[f(t)]$ and covariance function $\rm{cov}(f(t), f(t')) = k(t, t') = \mathbb{E}[(f(t)-m(t))(f(t')-m(t')]$, where $\mathbb{E}[f(t)]$ is the expectation value of the function $f(t)$ and $k(t, t')$ is referred to as the kernel function. The covariance function is used to populate the covariance matrix, $K_{t, t'}$, effectively setting the dependence of each point in the time-series to each other. As such $k(t, t')$ describes the overall variability. GP regression gives some functional form to $k(t,t')$, which it can then fit in order to give a functional form that best describes the input time-series. Given an input time-series $y_{i}(t_{i})$, this assumes $y_{i}(t_{i}) = f(t_i)$, i.e that our data are representative of the underlying GP function. However, real data have noise, such that this is modified to $y_{i}(t_i) = f(t_{i}) + \sigma^{2}_i$, where $\sigma^2_i$ describes the contribution of noise to each data-point. Hence, in our regression model we modify the covariance matrix of the input data, such that $\rm{cov}(y_i(t_i)) = K_{t,t'} + \sigma^2_i I$, where $I$ is the identity matrix and we set $\sigma^2_i$ to the value given by the error on each data-point. This effectively weights the contribution of each input data-point to the final GP model (and corresponding prediction). We stress that this \emph{only} affects the data used to train the GPs, effectively broadening the posterior. The predicted light-curves are all noiseless.

Each light-curve is trained independently, giving separate GP models for each dataset. Some previous works have used multi-output GPs to simultaneously train multiple light-curves on a single kernel function \citep[e.g][]{Lewin22}. While this has the advantage that it can improve the GPs predictive power, since it can use information from one light-curve when there is a gap in another, it does make the strong assumption that all light-curves in the training set are well described by the same variability process. Physically, this implies the same origin. In our case the UVW2 and HX light-curves have fundamentally separate physical origins, and so attempting to improve the GP predictions through multi-output learning could accidentally introduce artificial coherent signals. Using independent models \citep[as in, e.g][]{Griffiths21, Wilkins19, Lewin23, Lewin24}, we ensure that any strongly coherent signals between the light-curves are truly present in the data.

During our analysis we find that the rational quadratic (RQ) kernel provides the best description of both the UVW2 and HX datasets. This is perhaps unsurprising, as the RQ kernel allows the GP models to vary over multiple time-scales, which is what you would physically expect for a system consisting of multiple potentially interacting flows (i.e the disc and corona). For completeness, we also test for other commonly used kernel functions, mainly the squared-exponential (SE) and Matern kernels \citep[see][for a detailed description of the kernel functions]{Wilkins19}, and show the resulting Fourier products in Appendix\,\ref{app:GP_kerns}, which are all consistent with the RQ kernel predictions with the exception of the highest frequency bin.

There is one significant caveat in our choice of the RQ kernel (and also the commonly used SE and Matern variants), in that it assumes a stationary process. This is, of course, not necessarily the case in our light-curves, especially considering the significant changes seen in the SED throughout the campaign (section\,\ref{sec:SED_results}). However, considering the strong agreement with a simple interpolation scheme (Appendix\,\ref{app:lin_interp}), and the excellent sampling in the data, we continue our analysis using the RQ kernel, but keeping in mind the variability may be more complex than predicted by the GP models.

We optimize the kernel hyperparamaters by minimizing the negative log marginal likelihood \citep[see e.g][]{Rasmussen06, Wilkins19, Lewin22}. To control for the possibility of the solution finding a local rather than global minimum we re-fit the hyperparameters multiple times, each time using a different set of randomly drawn initial parameter values. We give an overview of the resulting summary statistics, along with the comparisons to the other kernels in Appendix\,\ref{app:GP_kerns}.

Fig.\ref{fig:GP_LCs} shows the GP model light-curves for UVW2 (top) and HX (bottom) as the solid green line, with the shaded region indicating its $1\sigma$ dispersion. For the remaining analysis we draw 5000 realisations of each light-curve \rev{from the best fit GP parameters}, on an evenly sampled grid with cadence $\Delta t = 1$\,day. This slightly oversamples the average cadence of the campaign, which we compensate for by discarding the highest frequency bin used in our power- and phase-lag-spectra, such that we only consider bins below the native Nyquist frequency of the data. These light-curve realisations give 5000 UVW2-HX pairs, used in the cross-spectrum to give the phase lags and coherences. Each pair is drawn using a separate random-seed, in order to avoid artificial high-frequency coherent signals. \rev{We stress that this is \emph{not} the same as independent light-curve realisations of a given noise process (as e.g in \citealt{Timmer95}), as the GP gives realisations centred around a best fit solution to a specific light-curve.}

\subsection{Power Spectra}

\begin{figure*}
    \centering
    \includegraphics[width=\textwidth]{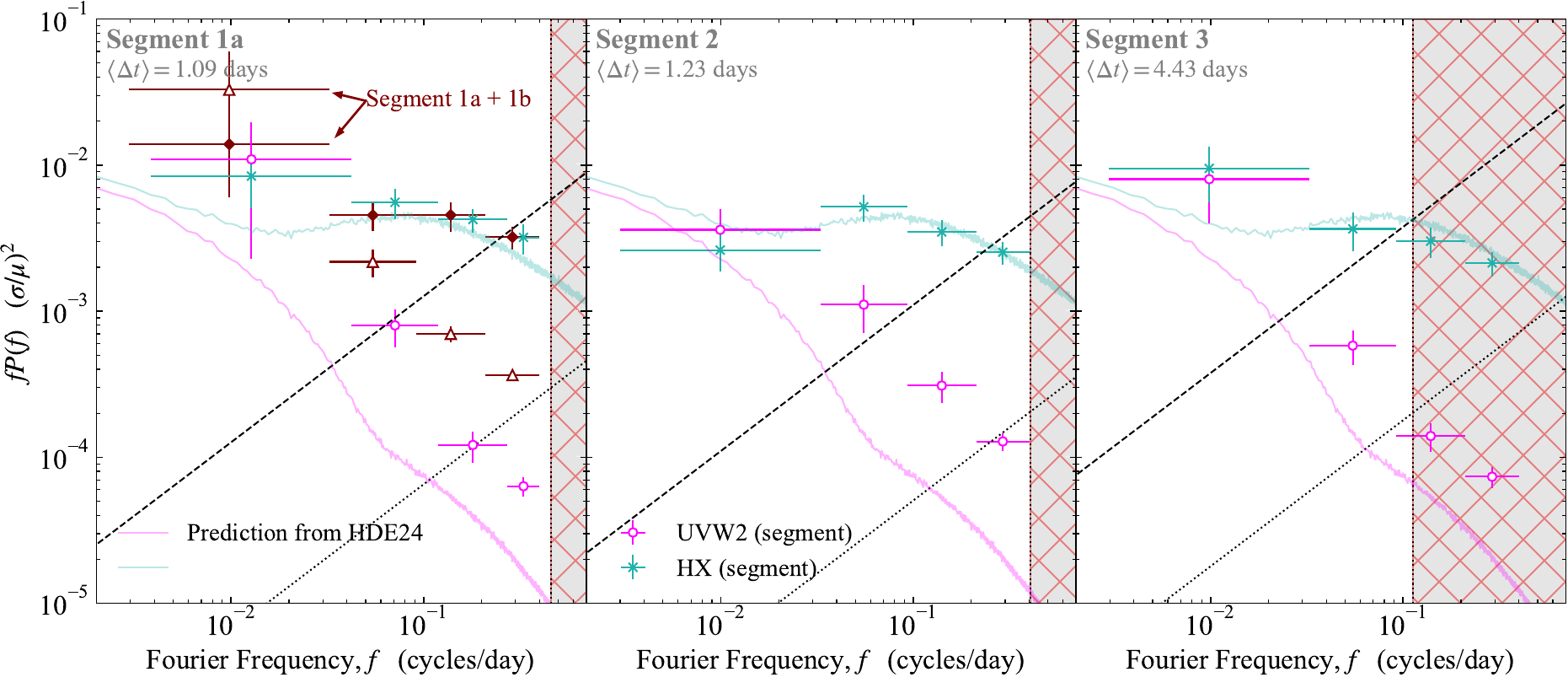}
    \caption{Power-spectra of Fairall\,9 for UVW2 (magenta open circles) and HX (turquoise stars) for each segment in the light-curves, as predicted by our Gaussian Process model. In the leftmost panel we also show the power-spectrum from Segment\,1a+1b as the open red triangles (UVW2) and filled red diamonds (HX). These have all been calculated by averaging over 5000 realisations of the GP model. The x-error indicates the width of the frequency bin, while the y-error \rev{is the combined error calculated from the standard error on the underlying noise process and the 1$\sigma$ dispersion from the 5000 GP realisations. At low frequencies the y-error is dominated by the noise intrinsic to the periodogram, while at high frequencies the dispersion in the GP realisations dominates.}
    The dashed black line shows the expected noise level for the HX power-spectra, while the dotted black line shows the noise for UVW2, both calculated following \citet{Uttley14}. We also include the predicted power-spectra from the propagating fluctuations model of \citetalias{Hagen24a} as the solid lines for both UVW (magenta) and HX (turquoise). The red hatched region displays the frequency range beyond $f_{NY}$ for each segment, calculated from the average sampling rate in each segment (given in the top left corner). Due to the significant drop in cadence in Segment\,3, the two highest frequency bins extend beyond $f_{NY}$, and are thus based on the properties learned by the GP model in previous (higher cadence) segments. As such they should not be trusted.}
    \label{fig:powSpec}
\end{figure*}

\begin{figure*}
    \centering
    \includegraphics[width=\textwidth]{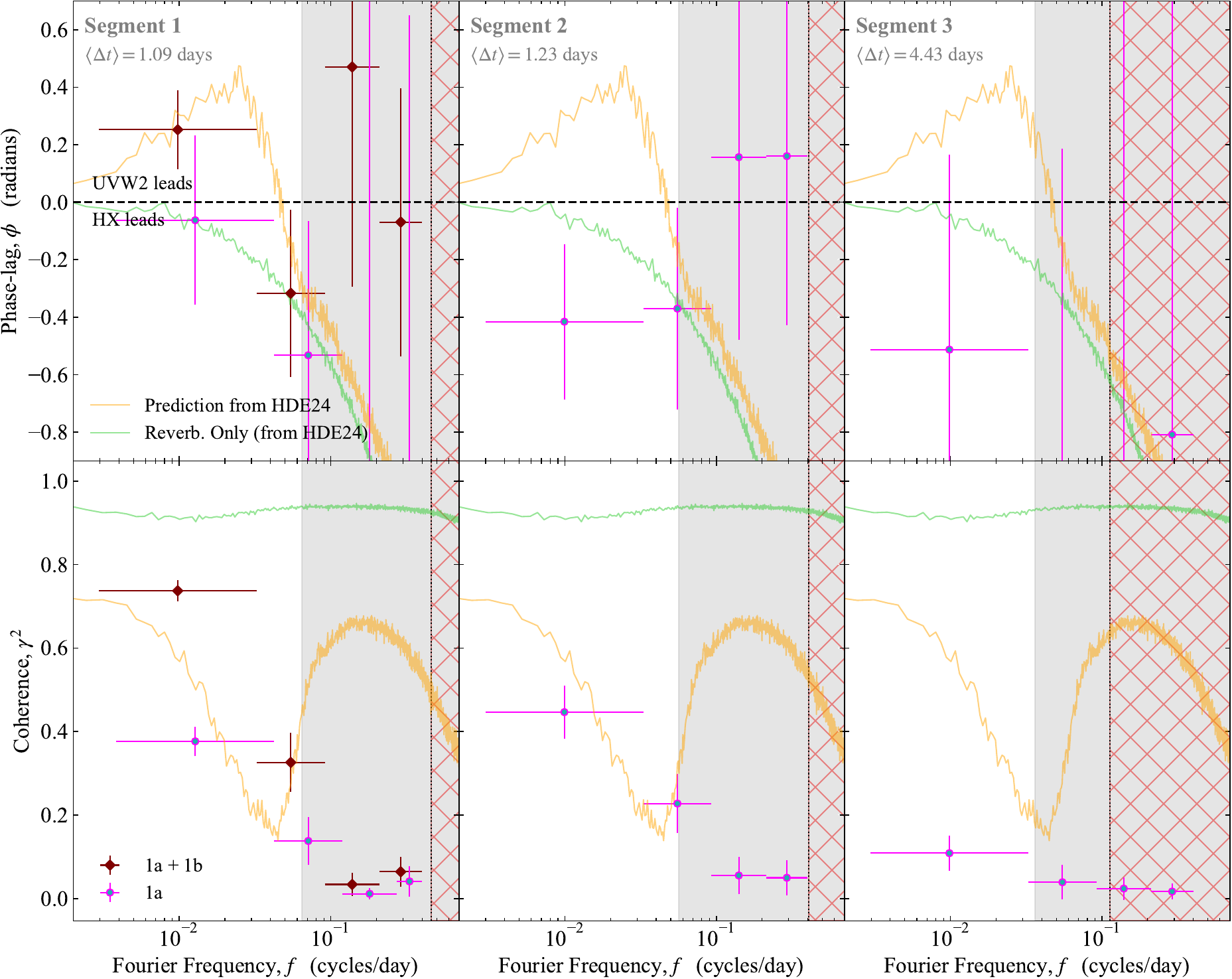}
    \caption{ \textbf{Top row:} Phase-lags \rev{for each segment} calculated between UVW2 and HX light-curves. These are defined such that a positive lag indicates UVW2 leads, while a negative lag indicates HX leads. \rev{These have been calculated by first generating cross-spectra from 5000 GP realisations, which are all binned in frequency. We then extract the phase-lag in each bin for each realisation before calculating the phase distribution in each bin from the 5000 realisation. The y-errors are then a combination of the intrinsic phase error (calculated from the coherence following \citealt{Uttley14} for the mean cross-spectrum) and the $1\sigma$ phase dispersion in each bin, added in quadrature.} \\
    \textbf{Bottom row:} Coherence calculated between the UVW2 and HX light-curves. \rev{Here the y-error represents the $1\sigma$ dispersion from 5000 GP realisations.} \\
    \textbf{Both rows:} In the leftmost column we also show the case for Segment\,1a+1b as the filled red diamonds. The solid lines show \emph{noiseless} model predictions for a combined propagation and reverberation scenario (from \citetalias{Hagen24a}, \textbf{orange}), and for a pure reverberation scenario (\textbf{green}, calculated using the model code and system parameters of \citetalias{Hagen24a}). The red hatched region displays the frequency range beyond $f_{NY}$ for each segment, calculated from the average sampling rate in each segment, while the grey region indicates where the phase and coherence become uncertain due to both gaps in the light-curve and the noise level in the PSDs (Fig.\,\ref{fig:powSpec} - \rev{see also Appendix\,\ref{app:GPsimulations}}), \rev{and so should not be trusted.}.
    }
    \label{fig:LagAndCoh}
\end{figure*}

We estimate the power spectral density (PSD) in each segment following \citet{Uttley14}. For any given light-curve $x(t)$ with corresponding Fourier transform $\tilde{X}(f)$, the periodogram is given by $|\tilde{X}(f)|^{2} = \tilde{X}^{*}(f)\tilde{X}(f)$ where the $^{*}$ denotes the complex-conjugate. Throughout this paper we normalise the periodogram by a factor $2N\Delta t / \mu$, where $N$ is the number of points in the time-series, $\Delta t$ is the cadence, and $\mu$ is the mean. This gives units of fractional variance per \rev{unit} frequency, such that the integral over the positive Fourier frequencies is approximately $(\sigma / \mu)^{2}$, where $\sigma$ is the variance in the light-curve.

The periodogram is only a random realisation of the underlying PSD \citep[see e.g][]{VanDerKlis89, Papadakis93, Timmer95}. To obtain an estimate of the PSD we need to average over multiple frequencies. Hence we define new frequency bins from $f_{\rm{min}} = 1/(N\Delta t)$ to $f_{\rm{max}} = f_{NY} = 1/(2\Delta t)$. We select bin edges that give integer number of Fourier frequencies within each bin. For the first bin (lowest frequency) we set this to $10$ frequencies, in order to give roughly Gaussian errors \citep{Epitropakis16}. For each subsequent bin, we double the number of frequencies, such that $N_{f, j} = 10\times 2j$, where $j$ is the bin-number, defined from $0$ for the low-frequency bin and $N_{f, j}$ is the number of frequencies in bin $j$. To avoid noise induced by the GP model on time-scales close to the Nyquist frequency we discard the highest frequency bin for all subsequent analysis. The resulting PSD, $P(f)$, is then the average of the periodogram contained within each bin, where we consider $f$ as the geometric centre of the bin. To ensure a robust interpretation, we have also performed this same analysis using synthetic light-curves, detailed in Appendix\,\ref{app:GPsimulations}. These generally show that given an input PSD, the GPs can confidently re-produce the low-frequency end of the power-spectrum, whereas the performance at high-frequencies depends somewhat on the input PSD with the GPs often overestimating the amount of power.

Fig.\ref{fig:powSpec} shows the resulting power-spectra for each segment (left to right) for both UVW2 (magenta hollow circles) and HX (turquoise stars). These have been generated by averaging over the PSD calculated for 5000 light-curve realisations from the GP model. Following \citet{Wilkins19, Lewin22}, the errors are given as the $1\,\sigma$ dispersion on the power between the separate GP realisations. We note that as the GPs are very well defined on long time-scales this likely leads to an underestimated error in the lower frequency bins.

In each segment the UVW2 power is clearly dropping off rapidly at high frequency, while the X-ray is relatively flat. While there is some uncertainty in the GP reconstructions in this frequency range (Appendix\,\ref{app:GPsimulations}), differences this large are real. It is challenging to smooth the X-ray signal on time-scales long enough to re-produce this through standard disc reverberation alone \citep[see e.g ][]{Arevalo08, Beard25}, suggesting instead two separate origins for the variability. 


At the lowest frequency bin there are strong variations in the power between each segment, after including the sharp rise at the end of Segment\,1a (giving Segment\,1a+1b). In some sense, this is expected, as this bin is only averaged over 10 Fourier frequencies, and so we expect it to be heavily affected by the stochasticity in the periodogram. What is interesting, however, is the fact that the variations in low-frequency X-ray power appears to follow that of the UVW2 power throughout (or vice-versa), even in the `stationary' segments excluding the rise. While this could be a fluke (there are only three segments after all), it could also be indicative of the UV X-ray relation. The seed-photons from the X-ray originate in the disc, and so if this is intrinsically variable at low frequency we would expect it to impact the low frequency X-ray also. If the UV goes up, then so must the X-ray, as the total X-ray emission is a sum of the power dissipated in the corona and the power originating from seed photons. \rev{There are some complications to this, as one also expects the spectral index to pivot. Depending on the observed band-pass, spectral pivoting could instead lead to an anti-correlation between the UV and X-ray. However, in Fig.\,\ref{fig:SEDcomp} it appears that our observed band-pass lies below the pivot point, such that we expect changes in the seed photon luminosity to lead to mostly correlated changes in the X-ray, unless there is interference from intrinsic X-ray variability.} \\

\subsection{Lags and Coherence}

The frequency resolved lags and coherence will give a better insight into the causal connection between the UV and X-ray. Following \citet{Uttley14}, the phase-lags are calculated from the cross-spectrum, $\tilde{C}(f) = \tilde{X}^{*}(f)\tilde{Y}(f)$, where $\tilde{X}$ and $\tilde{Y}$ are the Fourier transforms of two time-series $x$ and $y$ respectively. This is a complex number, given by $\tilde{C}(f) = |\tilde{X}| |\tilde{Y}| \exp(i\phi)$, where $\phi$ is the phase-difference between $\tilde{X}$ and $\tilde{Y}$. Physically, this corresponds to a lag between the Fourier components of $x$ and $y$ at a frequency $f$. 

As with the PSD, we need to bin the cross-spectrum to obtain meaningful results. Here we use the same frequency bins as in the PSD, and calculate the average cross-spectrum between each bin, $<C(f)>$. We also use the average cross-spectrum, along with the average PSDs, to obtain the coherence; $\gamma^{2} = <\tilde{C}(f)>^{*} <\tilde{C}(f)> / <P_{x}(f)> <P_{y}(f)>$. This is essentially a measure of the phase dispersion within each frequency bin \citep{Uttley14}\rev{, or more formally the fraction of the variability in two bands that can be related via a linear transform \citep{Vaughan97}}. If two time-series are well correlated over a given frequency range, then we expect a small phase-dispersion, giving a high coherence. \rev{Conversely}, for incoherent time-series, the phase-difference between Fourier components across a frequency bin will \rev{be highly variable}, giving low coherence. We show both the phase-lags and coherence in Fig.\ref{fig:LagAndCoh}.

The two highest frequency bins in each segment here are not to be trusted, as demonstrated by our simulations in Appendix\,\ref{app:GPsimulations}. The results in these bins are strongly affected by both gaps in the light-curve (indicated by the shaded grey region in Figs\,\ref{fig:LagAndCoh} and \ref{fig:toyMod_v_dat}) and the noise level (Fig.\,\ref{fig:powSpec}). Both of these effects compound to give a low (to non-existent) coherence, which in turn leads \rev{a poorly constrained lag}. Hence we will focus our analysis on the two lowest frequency bins in each segment. 
Though even here caution should be used, as while the GPs do a better job of reconstructing the input in this frequency range (Appendix\,\ref{app:GPsimulations}), there is still a systematic reduction in coherence (and thus averaged phase-lag) originating from GP uncertainties (Fig.\,\ref{app_fig:GPsim_constPhase}. Additionally, the low frequency bins are more susceptible to stochastic variations in the periodogram, due to averaging over a smaller number of frequencies. Nonetheless, Fig.\,\ref{fig:LagAndCoh} shows large differences in the lag and coherence, especially at low frequency, which is likely real.

Starting with Segments\,1a (i.e excluding the rise) and 2, we see very little change in both the phase-lags and coherence (Fig.\,\ref{fig:LagAndCoh}). The interesting part is the semi-coherent ($\gamma^2 \sim 0.4$) long low-frequency lag, with the UVW2 lagging the X-ray by $\phi \sim 0.4 - 0.75$\,radians ($=> \tau \sim 6 - 12$\,days at $f=10^{-2}$\,cycles/day). This could suggest reverberation, apart from the fact that the lag is far too long compared to what one expects for the light-travel time to the inner edge of the disc. Continuing to Segment\,3 we then see this low-frequency semi-coherent signal vanish. This is initially surprising, as while we do expect some reduction due to the sparser sampling in this section, the reduction seen here is considerably stronger than we see in our simulations (Appendix\,\ref{app:GPsimulations}). However, our simulations were performed for perfectly coherent input light-curves, and so perhaps this effect could be induced just through stochasticity; especially if there is some interference between the light-curves at this frequency. We will explore this more in the next subsection with our toy-model. 
\rev{Nonetheless, these results do appear to challenge the scenario suggested in \citetalias{Hagen24a}, where the UV X-ray connection is driven predominantly by propagating mass-accretion rate fluctuations. Though we do note that formally we cannot rule this out given the limitations of our data.}

Perhaps more challenging to reconcile with the drop in coherence between Segment\,1a/2 and 3 is the flip in lag when we include the rise at the end of 1a (Segment\,1a+1b; red diamonds in Fig\,\ref{fig:LagAndCoh}). This gives very clear, and highly coherent ($\gamma^2 \sim 0.75$), low frequency lead (i.e UVW2 leads HX) of $\phi \sim 0.35$\,radians ($=> \tau \sim 5.5$\,days at $f = 10^{-2}$\,cycles/day), which is certainly due to the sharp rise in the UVW2 light-curve. This corresponds to the portion of the campaign where the SED undergoes strong changes, clearly increasing the disc emission (Fig.\,\ref{fig:SEDcomp}), most easily explained with an increase in the mass-accretion rate, $\dot{m}$. This could then give a potential explanation for the flip from a lag to a lead. By definition, material moves from the outer to inner regions in accreting systems, and so we expect changes in $\dot{m}$ to occur in the outer regions \emph{first} and then move inwards. If this increase also leads to a more structural change, as suggested by the truncation radius in the SEDs, we would expect to see an effect in HX also, even though $\dot{m}$ fluctuations do not necessarily propagate into the corona itself, and so a highly coherent signal starting in the UV/optical and finishing in the X-ray.


The question then becomes what causes the tentative long low-frequency lag in Segments\,1a/2 and its subsequent disappearance (along with a drop in the coherence) in Segment\,3. If we assume for now that this long lag is real, then it corresponds to the X-rays leading the UV by $\phi \sim 0.4- 0.75$\,radians ($=> \tau \sim 6-12$\,days), which is significantly longer than the expected light-travel time to the disc. Instead, this suggests a large scale re-processor at BLR distances. However, this is problematic, as {\sc cloudy} simulations show the X-ray does not tend to re-process into the UV for BLR like densities \citepalias{Hagen24a}. Instead, the diffuse BLR emission should be induced by the EUV, which both carries most of the accretion power as well as a significant interaction cross-section. Additionally, the X-rays will see the low-frequency UV variability through the intercepting seed photons entering the corona. Hence, we suggest that perhaps this long low frequency reverberation lag, is in fact induced by the EUV illuminating the BLR, but because the same variability modulates the X-ray seed photons, which have a much shorter light-travel time, it appears like the X-ray lead.

This still does not explain the drop in coherence, and the lack of any coherent signal in Segment\,3. This could potentially be due interference, reducing the observed coherence. There is some evidence that the PSD intrinsic to the X-ray corona extends to lower frequencies \citep[e.g][\rev{Lefkir et al. in prep., private communication}]{Ashton22, Beard25}. In the case where this has of the order similar amplitude to the intrinsic UV PSD, it will contribute an incoherent signal to the low frequency bin which reduces the overall coherence. The exact amount of reduction depends strongly on the relative amount of incoherent to coherent variability power in the given frequency bin. If we then further speculate that the random nature of the periodogram, which recall is what we are actually measuring, could cause the relative amount of power in the incoherent to coherent variability to vary from segment to segment, we could in theory obtain the strong reduction in coherence seen in Segment\,3 along with the non-detection of any lag; as this requires coherent signals.

\subsection{Analytic Model - Definition}

In light of both the SED evolution and the changing lags, it is clear that the disc-corona connection evolves throughout the campaign. This is interesting, as Fairall\,9 is an `clean' AGN, and so it shows that even on moderately short time-scales (a few hundred days) the inner AGN structure is likely \rev{undergoes changes in geometry and structure}. However, it does introduce challenges in physically modelling the variability. In order to make a very rough estimate of the connection between the UV disc and X-ray corona during the three segments, we build a simplified analytic model. We start by assuming the disc and corona vary intrinsically as $L_{d}(t)$ and $L_{c}(t)$ respectively, and that these are the only two sources of intrinsic variability. We then aim to describe the observed UV and X-ray variability, $L_{uv}(t)$ and $L_{x}(t)$ respectively, as linear combinations of $L_{d}$ and $L_c$, with the form of the combination given by an input geometry. We give below a brief overview of the main components, and refer the reader to Appendix\,\ref{app:toy_mod} for a full derivation.

Starting with the X-ray, the observed emission will be a sum of the intrinsic coronal emission, $L_c$, and the incident seed-photon power $L_s$ \citep[e.g][]{Kubota18}. The seed-photons originate from the disc, and so these should vary as $L_d(t)$ but lagged and smoothed by the light-travel time $\tau_s$. As such we can write $L_s(t) = (L_d \circledast\psi_s)(t)$ where $\circledast$ denotes a convolution and $\psi_s(t, \tau_s, \delta t_s)$ is the \rev{impulse response function}, which we treat as a top-hat centred on a lag $\tau_s$ with width $\delta t_s$. This will give a light-curve lagged by $\sim \tau_s$ with respect to $L_d$ and with time-scales shorter than $\delta t_s$ smoothed. Note that though we treat $\delta t$ explicitly in the derivations, we will fix it at $\delta t = \tau$ in our analysis, as physically we expect the smoothing time-scale to be similar to the lag. If there is a propagation signal present, this will modulate $L_c$ with a component displaying variability like $L_d$, but lagged on a much longer time-scale \citep{Lyubarskii97, Arevalo06, Ingram11}. For simplicity we choose to ignore this, instead using $L_s$ as a proxy. If the seed-photon lag required to match the data is very long (i.e significantly more than expected from the inner edge of the disc), then it is likely that there is an additional slow moving inwards propagating signal present. Thus we have for the observed X-rays:

\begin{equation}
    L_x(t) = L_c(t) + (L_d \circledast\psi_s)(t)
\end{equation}

For the UV, this will be the sum of the intrinsic disc emission, $L_d$, a reverberation component originating from X-ray reprocessing, $L_r$, and a potential contamination from diffuse/line emission originating in a wind/the BLR, $L_w$. The reverberation component will naturally follow the X-ray light-curve. For simplicity we only allow it do depend on the intrinsic coronal variability, such that $L_r(t) = (L_c \circledast\psi_r)(t)$. Both diffuse free-bound and line emission in the wind/BLR will be induced by the EUV emission in the SED, and as such should have similar variability properties to $L_d$. Thus, we can write $L_w(t) = (L_d \circledast\psi_w)(t)$, such that:

\begin{equation}
    L_{uv}(t) = L_d(t) + (L_d \circledast\psi_w)(t) + (L_c \circledast \psi_r)(t)
\end{equation}

As well as encoding the lag and smoothing, the \rev{impulse response function} $\psi_y$, where $y=s,r,w$, also determine the amount of power, $\lambda_y$, transferred between each component. This sets the amplitude of the top-hat, such that if $\lambda_y = 0$ then there is no power-transferred by mechanism $y$ (i.e seed-photons, reverberation, or wind reprocessing), and if $\lambda_y =1$ then all the available power is transferred. In this formalism, $\lambda_y > 1$ implies we get more power out than we put in, which is unphysical.

The inclusion of a wind here is in part motivated by our previous work \citepalias{Hagen24a}, but mostly by the recent work of \citet{Partington24}. Their results showed that the hard X-ray emission leads the UV, but appears to follow the soft X-ray emission (see their Fig.\,6 - we do stress that the posterior on this measurement is rather broad, and so this is only tentative). This may seem initially counter-intuitive in an warm corona interpretation, however can be reconciled by the inclusion of the wind. The combination of disc photons entering the corona, modulating the X-ray, and re-processing off a wind, modulating the optical/UV, should impart a lagged coherent signal between the optical/UV and X-ray at low frequency. The sign of said lag will predominantly depend on the relative contribution and light-travel time for each component. If the wind contributes little to the overall emission, then the lag should be dominated by seed photons entering the corona, and thus the UV leads the X-ray. Conversely, if the wind is strongly prevalent in the overall emission, we should see the X-ray lead the UV; since the light-travel time for the seed photons is much shorter than that to the wind. 
\rev{A wind illuminated by EUV photons will predominately emit lines and free-bound diffuse emission, which does not extend into the soft X-ray band-pass. Hence in the soft X-rays one expects the lag to be entirely dominated by the seed-photons entering the corona}
Thus, one can relatively simply reconcile the phenomenology seen in \citet{Partington24} with the hard X-ray leading the UV but following the soft X-ray, while still having the UV and soft excess originate from the same component within the flow. A detailed study of this is beyond this paper, however we include a mention of it here as it acts as motivation for a portion of the model.

{\renewcommand{\arraystretch}{1.6}
\begin{table*}
    \centering
    \begin{tabular}{c c c c}
    \textbf{Component} & \textbf{Parameters (Unit)} & \textbf{Segment\,1a+1b} & \textbf{Segment\,1a, 2, 3} \\
    \hline 
    \hline 
    $|\tilde{L}_c|^2$ pars & $f_{b, x, \rm{min}}$ (cycles/day) & $10^{-3}$ & - \\
                           & $f_{b, x, \rm{max}}$ (cylces/day) & $4\times10^{-1}$ & - \\
                           & Norm. $(\sigma/\mu)^2$ & $3\times10^{-1}$ & - \\
    \hline
    $|\tilde{L}_d|^2$ pars & $f_{b, d}$ (cycles/day) & $10^{-4}$ & - \\
                           & Norm. $(\sigma/\mu)^2$ & $6 \times 10^{-1}$ & $2\times10^{-1}$ \\
    \hline
    $\psi_w$ (Wind)        & $\lambda_w$ & 0.3 & - \\
                           & $\tau_w$ (days) & 10 & - \\
    \hline
    $\psi_s$ (Seed-Photons) & $\lambda_s$ & 0.4 & 0.1 \\
                            & $\tau_s$ (days) & 10 & 1 \\
    \hline
    $\psi_r$ (Reverberation) & $\lambda_r$ & 0.2 & 0.1 \\
                             & $\tau_r$ & 1 & - \\
    \hline

    \end{tabular}
    \caption{Parameters for the Fairall\,9 analytic model. A $-$ in Segment\,1a,2,3 indicates it takes the same value as that in Segment\,1a+1b. We stress that this is \emph{not a fit}. Rather physically plausible parameters that could explain the data.}
    \label{tab:toyMod_pars}
\end{table*}
}

As we now have both $L_{uv}$ and $L_{x}$ in terms of $L_d$ and $L_c$, we can write their respective power- and cross-spectra as functions of $|\tilde{L}_d|^2$ and $|\tilde{L}_c|^2$, where $\tilde{L}$ denotes the Fourier transform of $L$. As we have also chosen simple top-hats for the \rev{impulse response functions}, $\psi_y$, then the power- and cross-spectra (and thus phase-lags and coherences) of $L_{uv}$ and $L_x$ are fully analytic, given an analytic description of $|\tilde{L}_d|^2$ and $|\tilde{L}_c|^2$ (see Appendix\,\ref{app:toy_mod}). Observations, as well as out light-curves, show UV/optical dominated by long term-variability, and strongly descending PSDs \citep{McHardy06, Edelson19, Beard25, Yu25}. Thus we describe $|\tilde{L}_d|^2$ as a zero-centred Lorentzian, with a characteristic break frequency of $f_{b, d} = 10^{-4}$\,cycles/day. This will approximate to a power-law like PSD, with $P(f) \propto f^{-2}$ over the frequency range covered by our data, and as such the choice of $f_b$ for the UV is rather arbitrary. For the X-ray there is \rev{extensive} evidence that the corona covers a range of variability frequencies, giving an intrinsically flat PSD \citep{Edelson99, Uttley02, Markowitz03, McHardy04, Ponti12a, Ashton22, Tortosa23, Beard25}. Hence, we choose to describe $|\tilde{L}_c|$ as a sum of equal amplitude zero-centred Lorentzian components, with break frequencies between $f_{b,x, \rm{min}}=10^{-3}$\,cycles/day to $f_{b,x,\rm{max}} = 4\times10^{-1}$\,cycles/day. For completeness, the intrinsic disc and coronal power-spectra we use are given by:

\begin{equation}
    |\tilde{L}_d|^{2}(f) = \frac{f_{b, d}}{(f_{b, d}^2 + f^2)}
\end{equation}

\begin{equation}
    |\tilde{L}_x|^2(f) = \sum\limits_{i} \frac{f_{b, x, i}}{(f_{b,x, i}^2 + f^2)}
\end{equation}

where $f_{b,x,i}$ is geometrically spaced from $f_{b. x. \rm{min}}$ to $f_{b,x,\rm{max}}$. These are then further re-normalised such that their integral over positive frequencies gives an input $(\sigma/\mu)^2$ (indicated by the Norm. parameter in table\,\ref{tab:toyMod_pars}).

This gives a fully analytic model, which simultaneously provides UV and X-ray power-spectra, the phase-lag, and the coherence.

\subsection{Analytic Model - Application}
\label{sec:toy_mod_app}

We now apply our analytic model to the Fairall\,9 data. The derivation given in Appendix\,\ref{app:toy_mod} gives the expected relation if averaged over infinite realisations (as we have intentionally averaged out the cross terms between $\tilde{L}_{uv}$ and $\tilde{L}_{x}$). However, as we measure only one realisations per segment, the data could easily deviate from the model prediction while still remaining consistent with a single realisation of the same model. To assess this we also evaluate the expected dispersion in each frequency bin by generating 10000 realisation of $\tilde{L}_{uv}$ and $\tilde{L}_x$, following the algorithm of \citet{Timmer95}, and then calculating the power- and cross-spectra directly from Eqns\,\ref{eqn:FT_Luv} and \ref{eqn:FT_Lx}.

\begin{figure*}
    \centering
    \includegraphics[width=\textwidth]{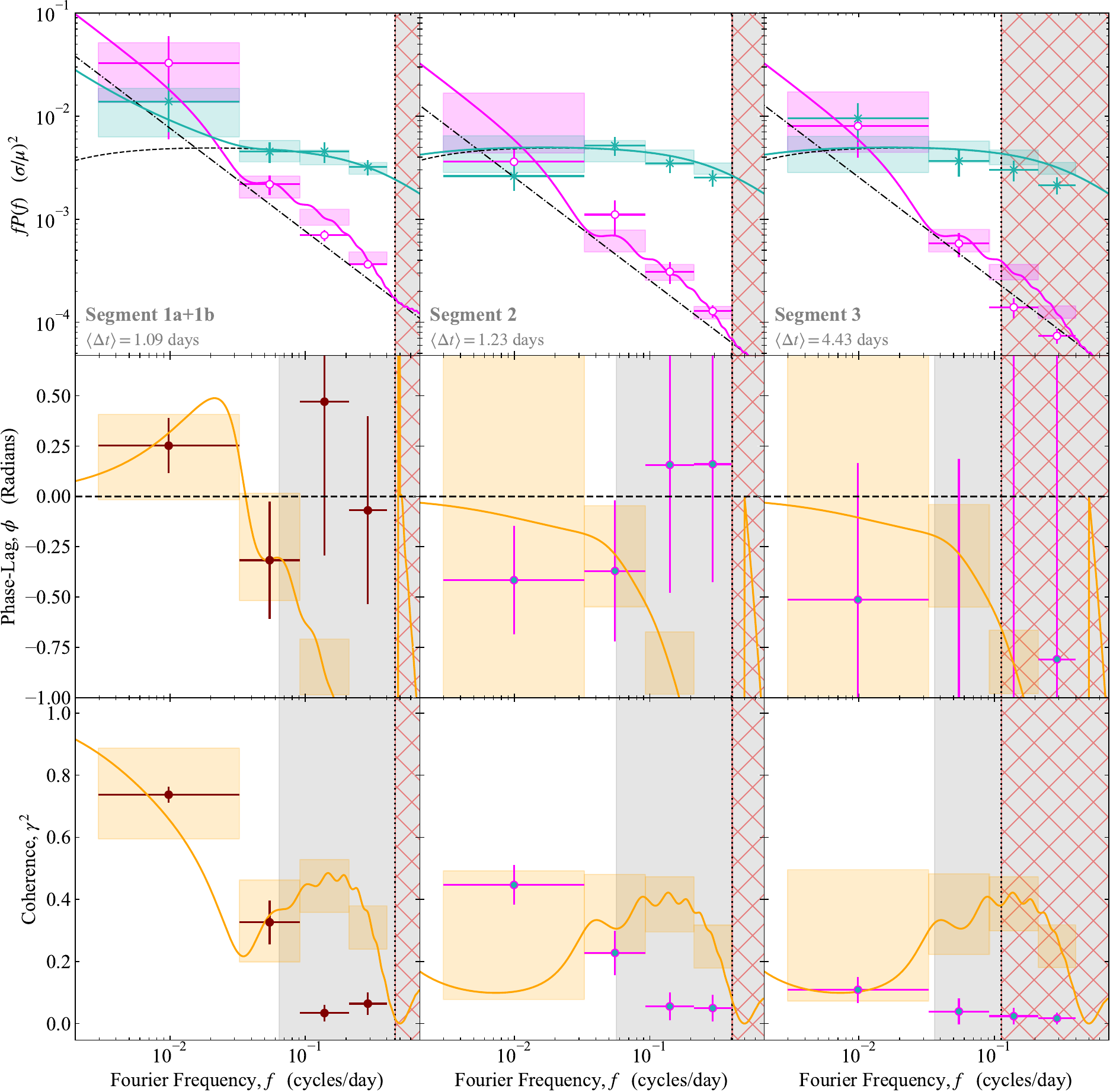}
    \caption{
    Analytic toy-model applied to the Fairall\,9 data. For clarity, we have elected not to show the Segment\,1a data, as this is \rev{mostly} consistent with Segment\,2 (Fig.\,\ref{fig:LagAndCoh}) and thus also the Segments\,2,3 model. \textbf{\textit{In all panels}} the solid coloured lines show the analytic model, which by definition is averaged over all realisations. The shaded squares show the expected $1\sigma$ dispersion in the model, calculated by evaluating the non-averaged model over 10000 realisations. The red hatched region indicate the frequency range beyond $f_{NY}$ for the given segment based off the average sampling in that segment, and the shaded grey region indicates the frequency range where the phase and coherence start to become uncertain due to the gap size and noise in the light-curves. The oscillatory behaviour in the model is an interference effect originating by adding together coherent, but slightly lagged, light-curves. We note that this is not a fit, due to the uncertainties in the data and simplicity of the model. Additionally, the model is \emph{noiseless}, and so will never recover the drop in high frequency coherence due to sampling and noise issues.  \\ \textbf{\textit{Top Row:}} Power-spectra. The solid lines show the total UV and X-ray model power-spectra (magenta and turquoise respectively), while the black dashed-dotted and dashed lines show the underlying disc and corona power-spectra respectively. The magenta open circles show the UVW2 data, while the turquoise stars show the HX data. \\ 
    \textbf{\textit{Middle Row:}} Phase-lags between UVW2 and HX, defined such that a positive lag implies UVW2 leads HX, and vice-versa for a negative lag. The solid orange line shows the analytic toy-model, while the circles show the data. \\
    \textbf{\textit{Bottom Row:}} Coherence between UVW2 and HX. As with the middle row, the solid orange line shows the analytic toy-model and the circles show the data. 
    }
    \label{fig:toyMod_v_dat}
\end{figure*}

We start by evaluating whether Segments\,1a/2 and 3 are consistent with being drawn from the same underlying model, as given the small differences in their SEDs we do not expect significant systematic differences in their accretion structure. The seed photons entering the corona should be dominated by emission from the inner disc \citep[e.g][]{Kubota18}. Likewise, the reverberation signal will also be dominated by the inner disc \citep{Hagen23a, Kammoun21a}. Hence we set their time-lag equal, such that $\tau_r = \tau_s$. The actual reverberation lag is slightly longer than the light-travel time to the inner edge of the disc, \rev{since the broad-band emission at each radius leads to a dilution of the reverberation signal}. Hence we fix $\tau_r = 1$\,day throughout. The wind we place at a time-lag of $\tau = 10$\,days, as this gives a distance of roughly $800-900\,R_G$, to remain consistent with what was required to match the cross-correlation lags in \citetalias{Hagen24a}.
For the underlying power-spectra we start by setting the amplitude of $|\tilde{L}_{c}|^2$ such that $|L_x|^2$ roughly matches the three high frequency bins in both Segments\,2 and 3, as these should be mostly unaffected by the seed-photon contribution. The low frequency bin should be strongly affected by the incident seed-photons, depending on the value of $\lambda_s$. For $|\tilde{L}_{d}|^2$, this will appear as a power-law in our frequency range (see dashed-dotted line in Fig.\,\ref{fig:toyMod_v_dat}). The additional contribution from reverberation to $|\tilde{L}_{uv}|^2$ will give a boost to the high-frequency end\rev{, as suggested in the recent sample of \citet{Yu25}}. The wind should only affect the low frequency bin, as due to the long light-travel time, variability for $f \gtrsim 1/(2\pi\tau_w) \sim 0.016$\,cycles/day will be strongly suppressed in the resulting wind PSD due to smoothing effects. Thus we now simultaneously adjust the amplitude of $|\tilde{L}_{d}|^2$, along with $\lambda_w$, $\lambda_s$, $\lambda_r$ to give a model that simultaneously matches the power-spectra, phase-lags and coherences in Segments 2 and 3. As we do not expect any propagation signal here, we have also fixed the seed-photon travel time to that of the reverberating photons, such that $\tau_s = \tau_r = 1$\,day. We note that for the coherence and phase-lags we focus on matching the two lowest frequency bins, as the high frequency bins are highly uncertain (see Appendix\,\ref{app:GPsimulations}).

\begin{figure*}
    \centering
    \includegraphics[width=\textwidth]{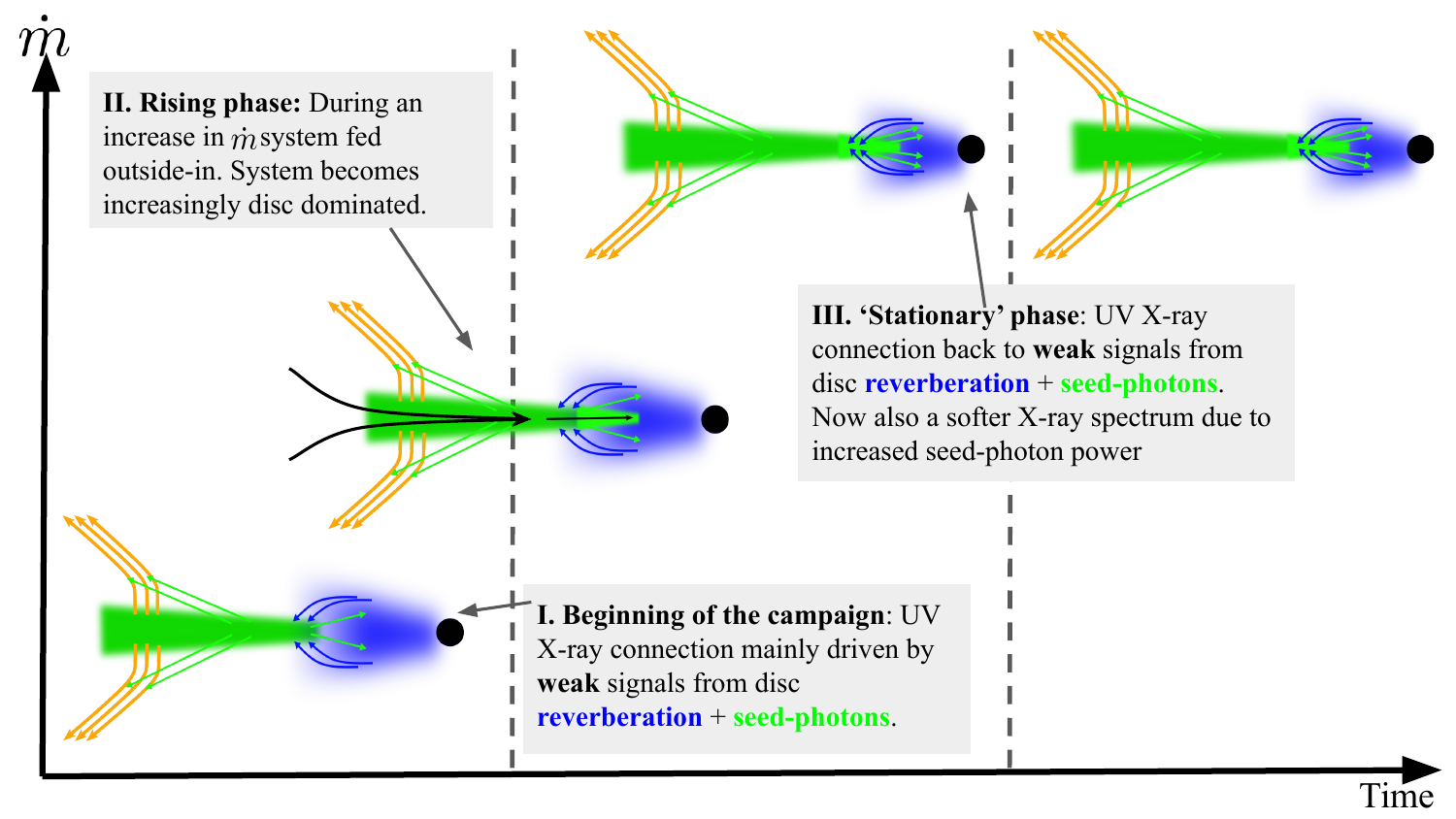}
    \caption{Cartoon depicting the UV-Xray relation throughout the Fairall\,9 campaign. Between segments\,1 and 2 there is an increase in $\dot{m}$. This is affects the system from the outside-in, giving a low-frequency lag where the UV leads the X-ray, as well as a significant softening of the X-ray spectrum due to an increase in UV seed-photons (see the SEDs in \ref{fig:SEDcomp}). From segment\,2 and onwards the system settles back into something resembling steady-state, with no propagation detected. Instead the variability is driven by intrinsic fluctuations from the disc and corona, which link directly to one another through seed-photons and disc re-processing. There is also an additional component originating in large-scale height material, either the BLR/wind, which re-processes the EUV emission, giving a long low-frequency reverberation lag. If the intrinsic X-ray PSD extends to lower-frequencies, then incoherent signals at low-frequency will at times wipe out the coherent signals, as the periodogram is only a random realisation of the underlying PSD, making the detection of a coherent lag essentially stochastic (i.e the main difference between segments 2 and 3).}
    \label{fig:cartoon}
\end{figure*}

This gives a model that can match both Segments\,1a/2 and 3, and is shown in Fig.\,\ref{fig:toyMod_v_dat} with corresponding parameters given in table\,\ref{tab:toyMod_pars}. The key here is that although the ensemble averaged model matches neither particularly well, the expected $1\sigma $ dispersion of the model \emph{is} consistent. What this shows is that \emph{if} the underlying coronal PSD has a similar amplitude to the underlying disc PSD at low frequencies, then it will contribute an interfering term to the total variability. For some realisations of the periodogram this will be stronger than others, giving varying coherences. In some sense, this can be seen visually in the light-curves in Fig.\,\ref{fig:GP_LCs}. Upon inspection it can be seen that the rise-and-plateau feature at the start of Segment\,3 (which will contribute most of the low-frequency power) in UVW2 appears to be partially present in the HX light-curve. But because there is an additional feature in the HX light-curve at the same-time, causing a drop, the overall coherence must go down. These competing signals at low-frequency will then subsequently imply that any lag measurement will likely vary depending on the underlying realisation, as the low coherence implies a wide phase-dispersion, hence giving the wide range in possible phase-lags seen in Fig.\,\ref{fig:toyMod_v_dat}. This also gives a possible explanation as to why the lag analysis of \citet{Edelson24} showed such a strong variation in amplitude (and also sign) throughout the campaign.

However, this does not simultaneously give predictions consistent with Segment\,1a+1b (though we stress only on the $1\sigma$ level), the main issues driving the disparity being the high low frequency coherence and strong boost to both X-ray and UV low-frequency PSDs. Matching the UV and x-ray PSDs requires a slightly higher amplitude in $|\tilde{L}_{d}|^2$, as well as a larger fraction of disc power entering the corona as seed-photons, $\lambda_s$, in order to give the low-frequency up-turn seen in the X-ray. This required increase in $\lambda_s$ is convenient, as it naturally also gives an increase in the overall coherence. But this then poses a separate problem, in that the increased coherence significantly narrows the range of possible phase-lags, which are now inconsistent with Segment\,1b. Hence, we also need to increase $\tau_s$ in order to recover the positive low-frequency lag seen here. In fact we require a rather long, unconstrained, lag of $\tau_s \gtrsim 8$\,days (note that in Fig.\,\ref{fig:toyMod_v_dat} we are showing the case for 10\,days, as though 8 is consistent it is marginal). This is much longer than expected from the light-travel time from the inner edge of the disc. It is also likely due to the sharp rise at in the light-curves at the end of Segment\,1b (Fig.\,\ref{fig:GP_LCs}), which contributes most of the signal at this frequency. It is perhaps not incidental then, that this coincides with an increase in the global mass-accretion rate, as seen by the SEDs (Fig.\,\ref{fig:SEDcomp}). This could then explain the coherent ingoing signal, as increases in mass-accretion should occur from the outer flow moving inwards. However, the SEDs also show that the overall X-ray power does not change (rather only the photon index). And so it is unclear whether the long inwards lag is due to increased seed-photons from the outer flow, fluctuations from the disc propagating into the corona, or a more structural change of the inner flow and thus UV to X-ray connection.

Finally, we stress here that though we have chosen to build an interpretation based off our analytic model, this does not correspond to a significant detection. The inconsistency in the model prediction of Segments\,1a,2,3 and the data in Segment\,1a+1b are only significant for the $1\sigma$ model dispersion. If we increase this to $2\sigma$, then we can mostly cover all segments with the same model (apart from the low frequency coherence in Segment\,1a+1b, which is still slightly off). Hence we stress that the interpretation of our timing analysis should be treated with some caution. However, due to the changes we do see being coincident with changes in the SED, which are highly significant at $>3\,\sigma$, we find it likely that the differences in the timing signatures between Segments\,1a+1b and 1a,2,3 are likely real, and represent changes to the accretion structure itself.

\section{Possible interpretation}

Given the change in SED shape throughout the campaign (section\,\ref{sec:SED_results}) we suspect the system has had an increase in the mass accretion rate, leading to potential changes in the inner structure. This rising phase aside, the timing-signatures between the UV and X-ray suggest that both the disc and corona are intrinsically variable, and only weakly correlated due to a combination of disc seed-photons entering the corona, and X-ray photons re-processing off the disc.

To aid the readers interpretation of the data and our model, we give in Fig.\,\ref{fig:cartoon} a cartoon picture of how we envisage the system behaves throughout the campaign. The key points to our interpretation (with the numerals corresponding to those in Fig.\,\ref{fig:cartoon}) are:

\begin{description}
    \item[\textbf{I. \textit{The beginning of the campaign:}}] The disc and corona vary independently, with the main link between the UV and X-ray driven by seed-photons from the disc entering the corona, and X-rays re-processing from the disc. There is also a wind, which re-processes the EUV, giving the long time-lags seen in \citet{Hernandez20}.

    \item[\textbf{II. \textit{The rising phase:}}] This corresponds to the steep brightening of the UVW2 light-curve, occurring between $\sim 58500$\,MJD and $\sim 58600$\,MJD (Fig.\,\ref{fig:GP_LCs}). We interpret this as a time where the mass-accretion rate is increasing, as suggested by the SEDs (Fig.\,\ref{fig:SEDcomp}). Assuming the system is fed from the outside-in, the this should imprint a long propagation lag in the optical/UV \citep[as seen in][]{Yao23}, which could also then give a propagation lag between the UV and X-ray, either through the increase in seed photons from the outer disc or by material from the disc propagating into the corona. We suspect this is likely driven by an increase in seed-photons, mainly due to the clear softening of the X-ray portion of the SED after this phase.

    \item[\textbf{III. \textit{The `stationary' phase:}}] The final two years of the campaign are characterised by highly similar SEDs, indicating the same accretion state, and are consistent with having the \emph{same} underlying PSDs. The strong changes in coherence and phase-lag at low frequency are then explained as being due to interference between the underlying disc and coronal variability. If the intrinsic corona PSD extends to lower frequencies, as suggested by some studies \citep{Ashton22, Beard25}, then this will inject an incoherent term in the cross-spectrum between the observed UV and X-ray lightcurves at low-frequency. What we measure is the periodogram, which is only a random realisation of the PSD, and so the relative strength of this incoherent term in the cross-spectrum can also vary; which will imprint a stochasticity into the coherence and subsequent lag. Given this, we have showed in section\,\ref{sec:toy_mod_app} that Segments\,2 and 3 are consistent with the same underlying model, even though an initial inspection of the data may suggest the contrary.
\end{description}

The key part of the UV and X-ray connection is the expected interference between the intrinsic disc and coronal emission. If these both had PSDs that were well separated in temporal frequency, then it becomes challenging to reproduce the moderate to low coherence. However, if the X-ray PSD intrinsically extends to lower frequency, as may be expected from an extended flow, then signals on similar time-scales will compete, reducing the overall coherence. For example if the UV goes up the seed-photons entering the corona must also go up, and so the X-rays go up. However, if the X-rays also have intrinsic variability on similar time-scales, they may go down, leading to a \emph{total} light-curve that appears incoherent from the UV. 

\rev{There are additional complications that could lead to reduced coherence between the UV and X-ray light-curves. Specifically, we expect differences in heating via fluctuations intrinsic to the corona and cooling from the disc seed photons to lead to a spectral pivoting behaviour in the X-ray. It has been pointed out by \citet{Uttley25} that these differences in heating and cooling of the corona can lead to complex lags and low coherences. The model of \citetalias{Hagen24a} does, however, include this effect, and still appears to overestimate the coherence (Fig.\,\ref{fig:LagAndCoh}), though we note that this is based off a single model rather than a proper fit to the data. Overall, while there are additional effects not accounted for in this paper, we currently suggest that to leading order the moderate to low observed coherence is likely due to interference between the intrinsic coronal and disc PSDs.}



\section{Conclusions}
\label{sec:conc}

The intensive monitoring campaigns, originally motivated by reverberation mapping, have highlighted the complex nature of AGN variability. Fairall\,9s clean view into the central engine, along with tis $\sim 1000$\,days of monitoring data \citep{Hernandez20, Edelson24, Partington24} make this one of the best suited targets for a detailed study of AGN variability processes.
We started this paper by examining the evolution of the SED throughout the campaign. This showed a strong change in the optical/UV disc component, likely due to a change in the mass-accretion rate, but very little change in the overall coronal X-ray emission other than a softening due to an increased contribution from disc seed photons.

Attempting to further resolve the UV-X-ray connection through a Fourier resolved timing analysis instead shows increased complexity and disparate lags. The most alarming, and challenging to reconcile, is the systematic drop in coherence between Segment\,1a+1b and Segments\,2 and 3. We attempt to address this with our analytic model, describing each observed light-curve as a linear combination of an underlying disc and coronal light-curve, which contains both direct emission as well as re-processed emission from a disc and wind and seed photon modulation of the X-ray. While it is possible to reconcile all Segments of the lightcurves as simply being random realisations of the same underlying, stochastic, model, this only works for model dispersions $>2\,\sigma$. Hence, we instead prefer the scenario where Segments\,2 and 3 are consistent with the same underlying model, where the UV and X-ray are only linked by disc reverberation and seed-photons travelling on the light-travel time; but then in Segment\,1 there is a more systematic change in the system giving an additional slow inwards moving component, which also imprints a long coherent low frequency lag with the UV leading. Given the systematic change in the SED, this seems more plausible.

The implications of the changing SED in the rising phase are that the systems \rev{inner structure evolves} on observable time-scales. The question then, which we will try to address in future work, is how does the accretion system respond on these time-scales. Results from both changing-look AGN \citep{Noda18, Temple23a, Panda24} and population studies \citep{Hagen24b, Kang25} suggest that the structure of the flow is strongly dependent on the (mass-scaled) mass-accretion rate. This then has implications for future timing-analysis, especially in the Fourier domain. As the signals we are measuring are fundamentally stochastic, one would ideally be able to average over many light-curve segments in order to properly constrain the real underlying lags and power \citep{Uttley14}. While this is commonly done in black hole \rev{X-ray} binary systems, as of yet it is challenging to do in AGN due to the long time-scales in the optical/UV (with the exception of studies exploring their very fast variability in the X-ray, \citealt{Kara13b, Kara16}). With the onset of LSST, and the current intensive monitoring data, one might think that we are entering an era where we can begin to average over AGN light-curve segments to give robust constraints on their timing signatures. However, if AGN systems \rev{change their inner structure} on even short time-scales, as suggested here, then that undermines this, \rev{since the use of ensemble averaging is only really meaningful if the system structure is mostly uniform between segments}. Future works should look to the evolution in the SED, and thus structure, \emph{as well} as their standard timing analyses to ensure physically meaningful results.


\rev{The rising SED, if interpreted as an evolution in the inner structure,} also brings about the more fundamental question of how can the system evolve on such short time-scales. \rev{The viscous time-scale for a standard \citet{Shakura73} disc is orders of magnitude too long compared to the time-scales seen in our data. 
This is a well known issue, which can be addressed by considering a modified disc structure, as this could allow for the disc to change its structure on shorter time-scales. In this framework, we speculate that the origin of the inner structural changes may be caused by either a change in the global mass accretion rate, propagating inwards from the outer regions, or by an instability in the disc itself closer to the inner regions.
Regardless of its origin, an evolution in the inner structure does suggest something fundamental about the nature of the flow, and would be avenue for our future work addressing and modelling the short time-scale structural variability in AGN}

\section*{Acknowledgements}

\rev{We would like to thank the anonymous referee for their constructive comments, which improved the quality of the manuscript.} 
We would also like to thank Matthew Temple for helpful conversations and feedback during the preparation of this manuscript.

SH acknowledges support from the Science and Technologies Facilities Council (STFC) through the studentship grant ST/W507428/1, as well as Barbera and Richard Ellis for the travel grant which facilitated the collaborations behind this paper.
CD acknowledges support from STFC through grant ST/T000244/1.

This work used the DiRAC@Durham facility managed by the Institute for Computational Cosmology on behalf of the STFC DiRAC HPC Facility (www.dirac.ac.uk). The equipment was funded by BEIS capital funding via STFC capital grants ST/K00042X/1, ST/P002293/1, ST/R002371/1 and ST/S002502/1, Durham University and STFC operations grant ST/R000832/1. DiRAC is part of the National e-Infrastructure.

This work made use of the following {\sc python} modules: {\sc numpy} \citep{Harris20}, {\sc GPy} \citep{GPy14}, and {\sc scipy} \citep{Virtanen20}. Additionally, the plots presented were made with both {\sc matplotlib} \citep{Hunter07} and {\sc seaborn} \citep{Waskom21}.

\section*{Data Availability}

The UVW2 and HX light-curves used here are from \citet{Edelson24}, and are available through the \swift\ archive (\url{https://www.swift.ac.uk/swift_live/index.php}). The data from the optical/UV data for the SEDs are from the same light-curves. The X-ray SED data, corresponding to the same light-curves, was built using the Leicester tool (\url{https://www.swift.ac.uk/user_objects/}) \citep{Evans09}.



\bibliographystyle{mnras}
\bibliography{Refs} 




\appendix

\section{Analytic Model Derivation}
\label{app:toy_mod}

We give here a detailed derivation of the analytic model used to interpret the power-spectra, phase-lags, and coherences for Fairall\,9. Throughout, we will use $\tilde{x}(f)$ to denote the Fourier transform of a time-series $x(t)$, and $f$ to represent temporal frequency. The Fourier transform results in a complex valued function, which we will represent throughout in polar form, such that for a complex number $c = |c| e^{i\phi} = |c|(\cos(\phi) + i\sin(\phi))$, which has a complex-conjugate $c^* = |c|e^{-i\phi}$. Finally, we use the following definition for the Fourier transform:

\begin{equation}
    \mathscr{F}[x(t)](f) = \int_{-\infty}^{\infty} x(t) e^{i2\pi t f} df
\end{equation}

This model assumes that both the observed UV and X-ray light-curves, $L_{\rm{uv}}(t)$ and $L_{x}(t)$ respectively, can be written as a linear combination of an intrinsic disc light-curve, $L_{d}(t)$, and an intrinsic coronal light-curve, $L_{c}(t)$.

The X-ray emitted power will be set by the intrinsic coronal power. Additionally, there will be a contribution from seed photons originating in the disc entering the corona, $L_s(t)$, such that $L_{x}(t) = L_{c}(t) + L_{s}(t)$. These seed-photons will have an identical light-curve as the intrinsic disc light-curve, but lagged and smoothed on some light-travel time $\tau_{s}$. We can therefore write $L_s(t) = (L_{d} \circledast \psi_{s})(t)$, where $\psi_{s}(t)$ is a transfer-function encoding the time-lag, smoothing, and relative power between $L_{d}(t)$ and $L_{s}(t)$, and $\circledast$ denotes a convolution. There can also be an additional contribution to $L_{x}(t)$ in the form of fluctuations from the disc propagating into the corona. These should also follow $L_{d}(t)$ in shape, but be lagged on considerably longer time-scales than $L_{s}$, and give a multiplicative rather than additive contribution \citep{Ingram11, Ingram12, Ingram13}. However, in the interest of simplicity we ignore this contribution, and rather state that if the required time-lag in $\psi_{s}(t)$ is much larger than the expected light-travel time, then it is likely that an additional inwards propagating signal is required.

The UV emission will consist of the intrinsic disc emission as well as a potential contribution from diffuse emission originating in the BLR/a wind \citep{Korista01, Korista19, Netzer22, Hagen24a}, $L_{w}(t)$. This diffuse emisison will predominantly be induced by the EUV disc emission, such that we can write $L_w(t) = (L_{d} \circledast \psi_w)(t)$, where $\psi_{w}$ encodes the light-travel time from the inner disc to the BLR/wind. Additionally, variable X-rays from the corona will be incident on the disc, driving a reverberation signal, $L_{r}(t)$. For simplicity, we consider this to follow $L_{c}(t)$ rather than $L_{x}(t)$, such that $L_{r}(t) = (L_{c} \circledast \psi_{r})(t)$. We then have for our total UV and X-ray light-curves:

\begin{align}
    L_{uv}(t) &= L_{d}(t) + (L_{d} \circledast \psi_{w})(t) + (L_{c} \circledast \psi_{r})(t) \\
    L_{x}(t) &= L_c(t) + (L_{d} \circledast \psi_{s})(t)
\end{align}

To calculate the power-spectra, phase-lags and coherence between $L_{x}$ and $L_{uv}$ we need to evaluate their Fourier transforms, $\tilde{L}_{uv}(f)$ and $\tilde{L}_{x}(f)$. Using the linearity properties of the Fourier transform, combined with the convolution theorem, this gives:

\begin{align}
    \label{eqn:FT_Luv}
    \tilde{L}_{uv}(f) &= \tilde{L}_{d}(f) + \tilde{L}_{d}(f) \tilde{\psi}_{w}(f) + \tilde{L}_{c}(f) \tilde{\psi}_{r}(f) \\
    \label{eqn:FT_Lx}
    \tilde{L}_{x}(f) &= \tilde{L}_{c}(f) + \tilde{L}_d(f) \tilde{\psi}_{s}(f)
\end{align}




Realistic \rev{impulse response functions} not only encode the lag between two light-curves, they also include relevant geometric considerations, response amplitudes, and relativistic effects \citep[e.g][]{Kammoun21b}. This is too complex for our model, and thus we instead opt for a simple top-hat function, as also done in \citet{Kawamura22}. We centre the top-hat on a characteristic lag $\tau$ and give it a width $\delta t$ which gives the smoothing time-scale. In our analysis we set $\delta t = \tau$, as physically the signal should to be smoothed on lag-time. For completeness, however, we include it explicitly throughout our derivation. Thus, our \rev{impulse response functions} are given by:

\begin{equation}
    \psi_{y}(t) =
    \begin{cases}
        \frac{\lambda_{y}}{\delta t_{y}} \,\, \rm{for} \,\, \tau_y-\frac{\delta t_y}{2} \leq t \leq \tau_{y} + \frac{\delta t_y}{2} \\
        0 \, \, \rm{else}
    \end{cases}
\end{equation}

Here $y=w, s, r$ is used as joint notion for the wind reprocessing, seed-photon propagation, and disc reverberation cases. $\lambda_y$ is a normalisation factor, that describes the fraction of power from the incident emission re-processed into the output emission. The main advantage of a top-hat \rev{impulse response function} is that its Fourier transform is analytic, giving \rev{the transfer function}:

\begin{align}
    \begin{split}
        \tilde{\psi}_y(f) &= \frac{\lambda_{y}}{\pi f \delta t_y} \sin(2\pi f \delta t_{y}) e^{i2\pi f \tau_y} \\
        &=  \Lambda_{y} \sin(\Phi_{y})e^{i\Theta_y}
    \end{split}
\end{align}

where $\Lambda_y(f) = \lambda_y/(\pi f \delta t_y)$, $\Phi_y(f) = 2\pi f \delta t_y$, and $\Theta_y(f) = 2\pi f \tau_y$ are used throughout to simplify the notation. We also note that hereinafter we drop the explicit reference to these being functions of $f$. 

While we derive the power- and cross-spectra we will often find ourselves evaluating the cross-terms between $L_{d}$ and $L_c$, given by:

\begin{align}
    \begin{split}
        \tilde{L}_d^* \tilde{L}_c &= |\tilde{L}_d||\tilde{L}_c| e^{i(\phi_c - \phi_d)} \\
        &= |\tilde{L}_d||\tilde{L}_c| e^{i\Delta \phi_{cd}}
    \end{split}
\end{align}

$\Delta \phi$ can only take values between $-\pi$ and $\pi$ \citep[e.g][]{Uttley14}. Assuming $L_c(t)$ and $L_d(t)$ are two incoherent time-series, then $\Delta \phi_{cd}$ should form a uniform distribution between $-\pi$ and $\pi$ when we average over an infinite number of realisations. The average of $e^{i\phi}$ when $\phi \sim \mathcal{U}(-\pi, \pi)$ is:

\begin{align}
    \begin{split}
        <e^{i\phi}> = \frac{1}{2\pi} \int_{-\pi}^{\pi} e^{i\phi} d\phi = 0
    \end{split}
\end{align}

Thus, on average we can write $\tilde{L}_d^* \tilde{L}_c = \tilde{L}_c^* \tilde{L}_d = 0$, eliminating these cross-terms throughout. This simplifies the maths, and allows us to write fully analytic expressions for the power- and cross-spectra. These are given by (using equations A4 and A5): \\

\noindent
\textbf{\textit{X-ray power:}}
\begin{align}
    \begin{split}
        |\tilde{L}_{x}|^2 &= |\tilde{L}_c|^2 + |\tilde{L}_d \tilde{\psi}_s|^2 \\
        &= |L_{c}|^2 \Lambda_s^2 \sin^2(\Phi_s)
    \end{split}
\end{align}
\\

\noindent
\textbf{\textit{UV power:}}
\begin{align}
    \begin{split}
        |\tilde{L}_{uv}|^2 &=|\tilde{L}_d|^2 + |\tilde{L}_d \tilde{\psi}_w|^2 + |\tilde{L}_c \tilde{\psi}_r| + |\tilde{L}_d|^2\left( \tilde{\psi}_w + \tilde{\psi}_w^* \right)  \\
        &= |\tilde{L}_d|^2 \left\{ 1 + \Lambda_w^2 \sin^2(\Phi_w) + 2\Lambda_w \sin(\Phi_w) \cos(\Theta_w) \right\} \\
        &\,\,\,\,\,\,\,\, + |\tilde{L}_c|^2 \Lambda_r^2 \sin^2(\Phi_r)
    \end{split}
\end{align}
\\

\noindent
\textbf{\textit{Cross-Spectrum:}}
\begin{align}
    \begin{split}
    |\tilde{C}| &= \tilde{L}_{uv}^* \tilde{L}_x = |\tilde{L}_d|^2 \left\{ \tilde{\psi}_s + \tilde{\psi}_w^* \tilde{\psi}_s \right\} + |\tilde{L}_c|^2 \tilde{\psi}_r^*  \\
    \end{split}
\end{align}

This has corresponding real and imaginary parts:

\begin{align}
    \begin{split}
        \mathfrak{Re}[\tilde{C}] = &|\tilde{L}_d|^2 \Lambda_s \sin(\Phi_s) \Big\{ \cos(\Theta_s) + \Lambda_w \sin(\Phi_w) \cos(\Theta_s - \Theta_w) \Big\} \\
        &+ |\tilde{L}_c|^2 \Lambda_r \sin(\Phi_r) \cos(\Theta_r)
    \end{split}
\end{align}

\begin{align}
    \begin{split}
        \mathfrak{Im}[\tilde{C}] = &|\tilde{L}_d|^2 \Lambda_s \sin(\Phi_s) \Big\{ \sin(\Theta_s) + \Lambda_w \sin(\Phi_w) \sin(\Theta_s - \Theta_w) \Big\} \\
        &- |\tilde{L}_c|^2 \Lambda_r \sin(\Phi_r) \sin(\Theta_r)
    \end{split}
\end{align}

This then gives the model phase-lag as $\phi = \arctan(\mathfrak{Im}[\tilde{C}]/\mathfrak{Re}[\tilde{C}])$. Additionally, since we are ensemble averaging over all realisations by cancelling the cross-terms, we can use the above expression to give an analytic expression for the coherence, $\gamma^{2} = (\tilde{C}^*\tilde{C})/(|\tilde{L}_{uv}|^2 |\tilde{L}_x|^2)$. We note that from our definition of the Fourier transform (A1) and cross-spectrum (A13), a positive phase-lag implies $L_{uv}(t)$ leads $L_{x}(t)$, while a negative lag implies the inverse.

\section{{\sc agnsed} Contour Plots}
\label{app:SED_contour}

\begin{figure*}
    \centering
    \includegraphics[width=\textwidth]{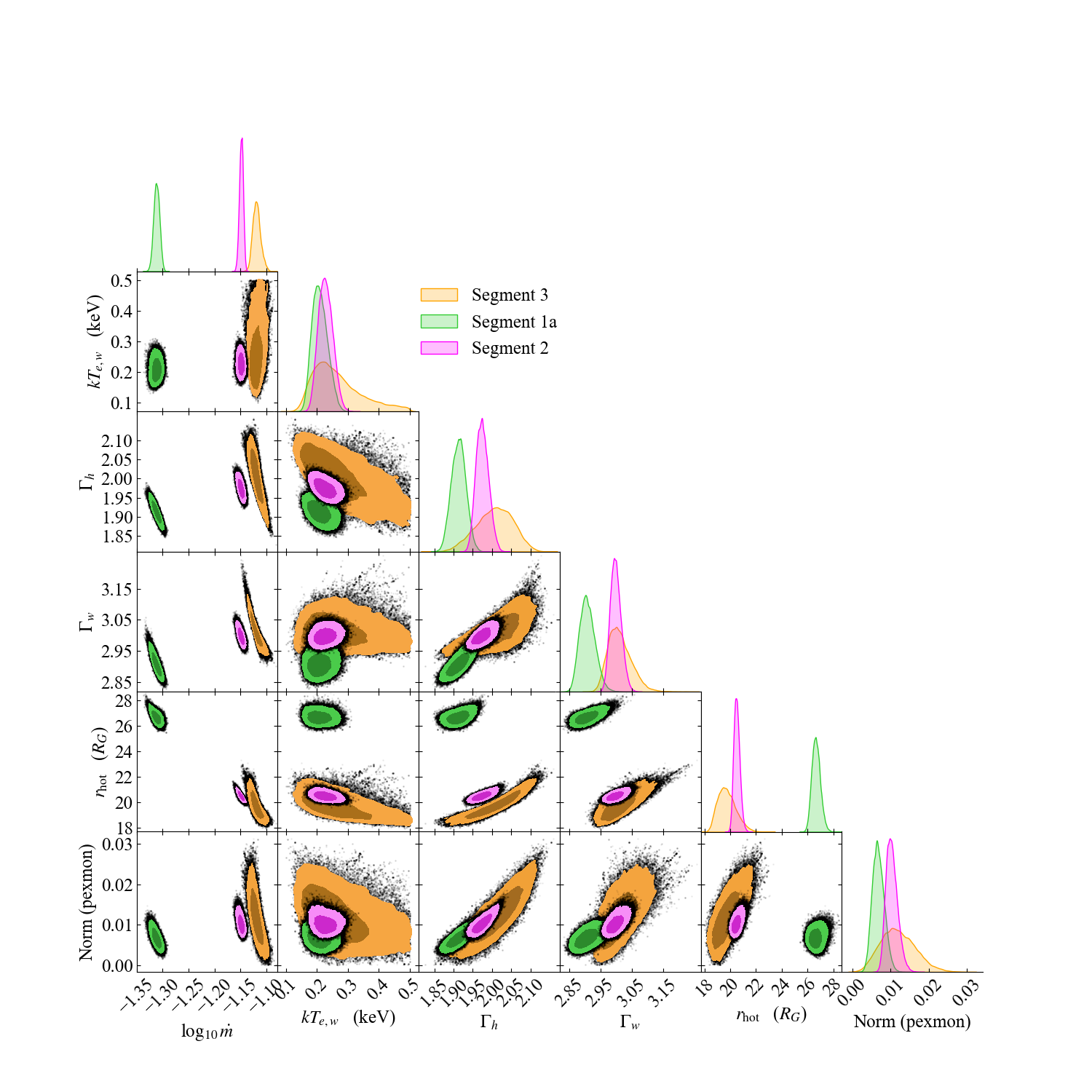}
    \caption{Contour plot for our MCMC runs for the SED on each segment, where segment 1, 2 and 3 are given by the green, magenta, and orange contours respectively. The levels correspond to $1\sigma$ and $2\sigma$ confidence. The mass-accretion rate is clearly distinguished between each segment. For the remaining parameters we can generally distinguish between segments 1 and 2 at high confidence. However, for segment 3 the size for the contours increases significantly due to lower signal-to-noise in the \swift-XRT spectra as there are fewer observations in the segment to average over.}
    \label{app_fig:SEDcontours}
\end{figure*}

\section{GP reconstructions from simulated data-products}
\label{app:GPsimulations}

\begin{figure*}
    \centering
    \includegraphics[width=\textwidth]{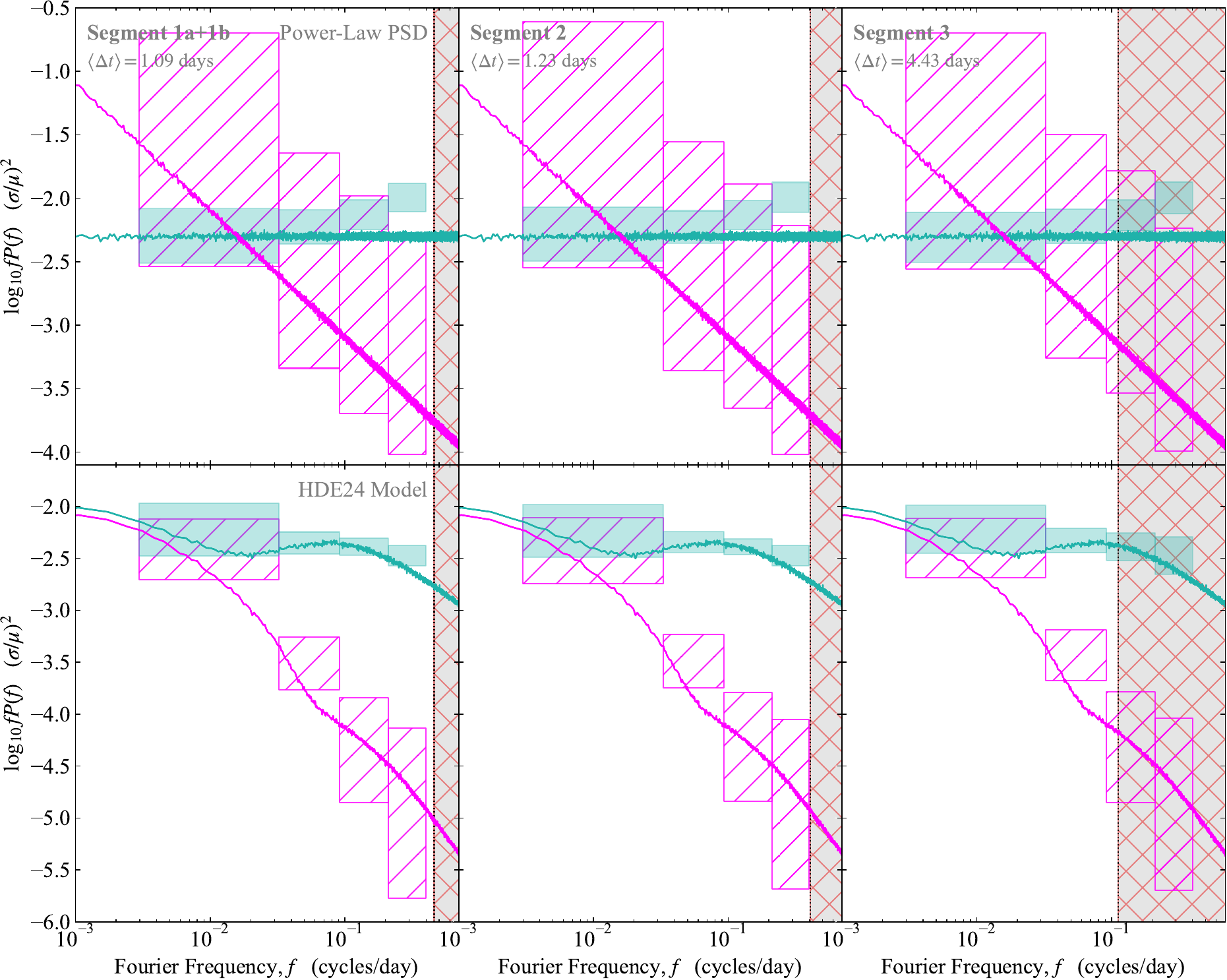}
    \caption{Reconstructed power-spectra from power-law PSDs (top) and model PSDs from \citetalias{Hagen24a} (bottom). the solid coloured lines indicate the input PSDs, while the boxes indicate the reconstruction. The width of the boxes gives the width of the frequency bins used in our analysis, and the height is the $1\sigma$ dispersion from $1000$ realisations. The segments correspond to those used in our main analysis}
    \label{app_fig:GP_SimPowSpec}
\end{figure*}

\begin{figure*}
    \centering
    \includegraphics[width=\textwidth]{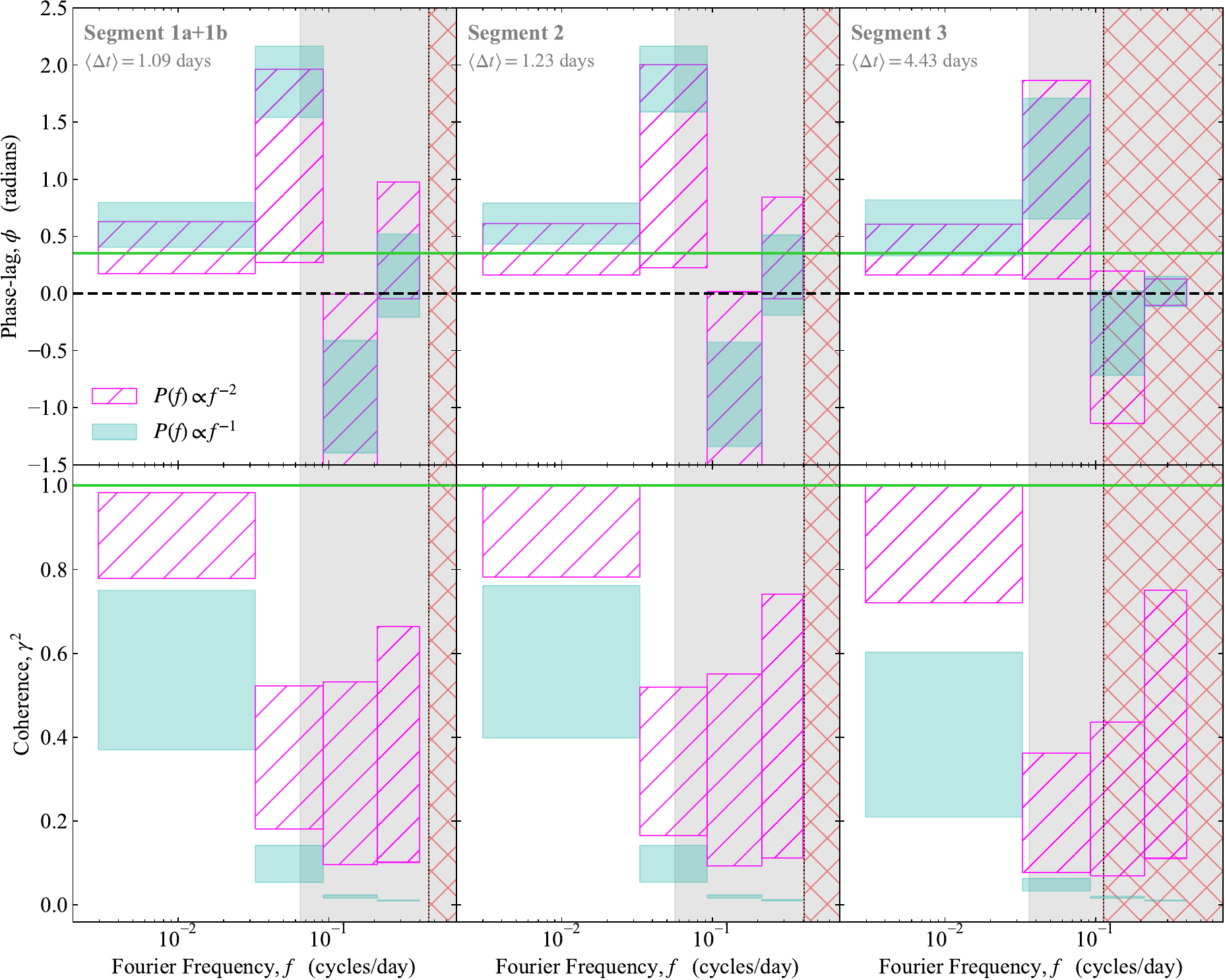}
    \caption{Reconstructed phase-lags (top) and coherences (bottom). Here we have input two originally identical light-curves, but lagged by a constant phase-lag of $\phi = 0.35$ and including a contribution from data-like noise. The height of the boxes give the $1\sigma$ dispersion, calculated from $1000$ realisation, while the width gives the width of the frequency bins. The GPs clearly have issues at higher frequency, where gaps in the light-curves give incoherent contributions due to the random nature of the GPs.}
    \label{app_fig:GPsim_constPhase}
\end{figure*}

In this appendix we test the GPs ability to re-construct light-curves with known power-spectra, coherence and phase-lags. We start by simulating pairs of light-curves, each offset by a constant phase-lag of $\phi=0.35$, to roughly match the low-frequency lags/leads we see in the real data, following the algorithm of \citet{Timmer95}. As such each light-curve pair is expected to have perfect coherence (i.e $\gamma^2 = 1$ for all $f$). 
\rev{To account for potential leakage effects \citep{Alston13}, due to the long time-scales present in the light-curves, we simulate each light-curve pair for a total duration of $15000$\,days at a cadence of $0.1$\,days, before re-sampling onto an uneven time-grid matching the real data. One matching the UVW2 data, and the other matchin the HX data.}
We then also add noise and errors to these, according to the real UVW2 and HX data respectively, creating data-like light-curves. Each data-like light-curve pair is then passed through the same pipeline as to our real data which: individually trains a GP with an RQ kernel on each light-curve, extracts $5000$ GP realisations for each light-curve, and then calculates power-spectra, phase-lags, and coherences in each bin.

The above process is repeated $1000$ times, for both in input PSD of the form $P(f) \propto f^{-2}$ (i.e optical/UV like) and $P(f) \propto f^{-1}$ (i.e X-ray like). We use both PSD types to give a rough assesses on how the different variability time-scales affects the GP predictions. In Figs.\,\ref{app_fig:GP_SimPowSpec} and \ref{app_fig:GPsim_constPhase} we show the $1\sigma$ dispersion in the resulting power-spectra, and phase-lags and coherences, respectively.

Starting with the power-spectra, at low-frequency the GPs generally \rev{re-produce the input PSD well}, for both $f^{-2}$ and $f^{-1}$ PSDs. At high frequency, however, there does appear to be an upturn, an effect particularly prominent in the $f^{-1}$ case. This is above the noise level expects for these errors and sampling. It is uncertain whether this is an aliasing effect, or an over prediction of the high frequency variability induced by the GP models. Either way, we treat the high-frequency data with caution in our main analysis, instead choosing to focus on low frequencies.

For the phase-lags and coherence we see a systematic reduction in both phase and coherence as we go to higher frequencies. This is not surprising. Firstly, noise starts to become important in the two high frequency bins, and so we expect some reduction in the coherence here. Secondly, in places where the light-curves have gaps, the GPs are simply `guessing'. While they will be informed by data on either side of said gap, for short time-scales (smaller than the gap length) they will be effectively random. The effect then, when comparing two GP predictions, is that we will have an overall reduction in coherence on time-scales near or faster than the typical gap length in the light-curve. This will naturally lead to a simultaneous reduction in the phase-lag, simply because two incoherent light-curves will have phase-lags uniformly distributed between $-\pi$ and $\pi$, and so when we average the phase in each frequency bin this tends to $0$. In terms of our analysis, this implies that we can only really trust the low frequency results.

\section{Fourier lags through linear interpolation}
\label{app:lin_interp}

\begin{figure*}
    \centering
    \includegraphics[width=\textwidth]{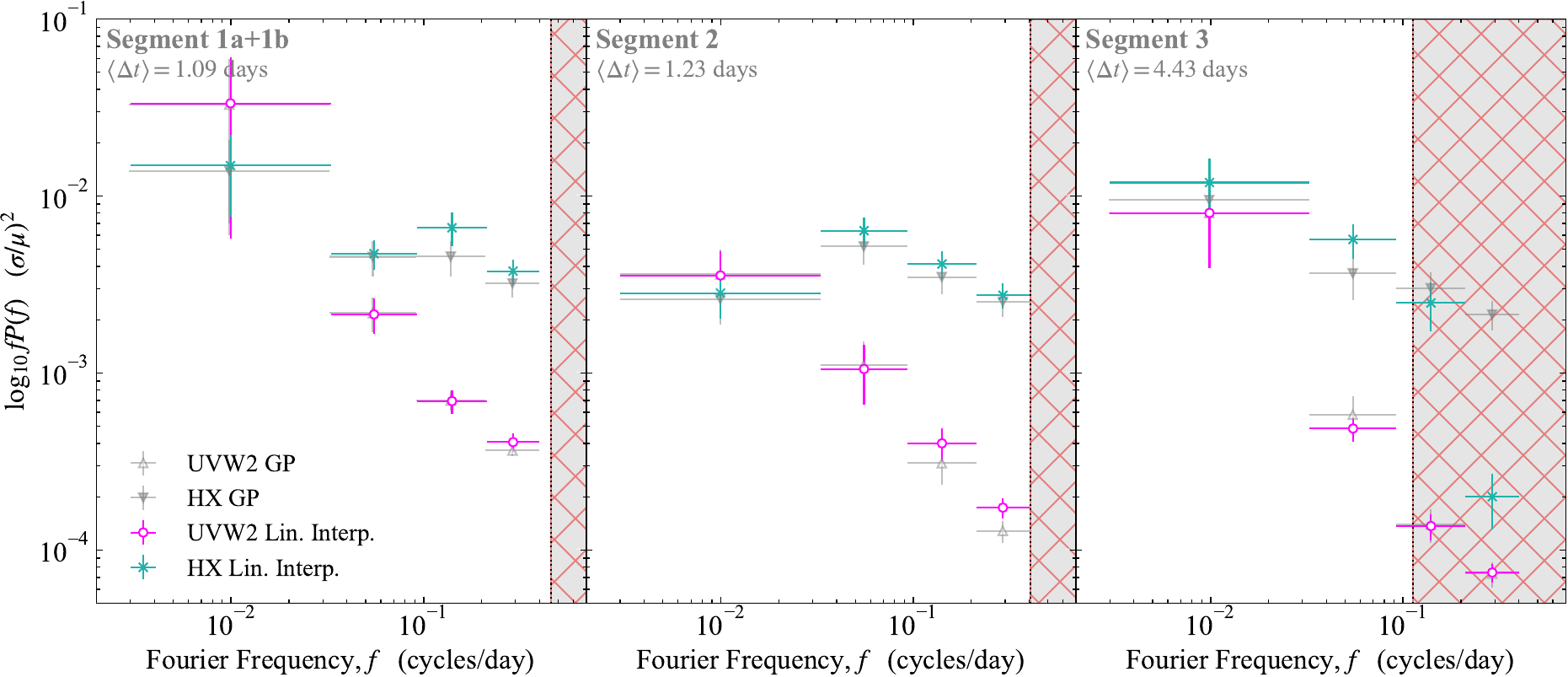}
    \caption{Comparison of the power-spectra predicted from our GPs (grey points) and the power-spectrum calculated by linearly interpolating the light-curves (magenta and turquoise for UVW2 and HX respectively). For Segments\,1a+1b and 2 there is very little difference between the two. For Segment\,3 we see an under prediction of the highest frequency HX variability through linear interpolation. This is simply due to the decreased sampling in Segment\,3, leading to these frequency range lying beyond the native Nyquist frequency, as indicated by the red hatch region.}
    \label{app_fig:powSpec_linInterp}
\end{figure*}

\begin{figure*}
    \centering
    \includegraphics[width=\textwidth]{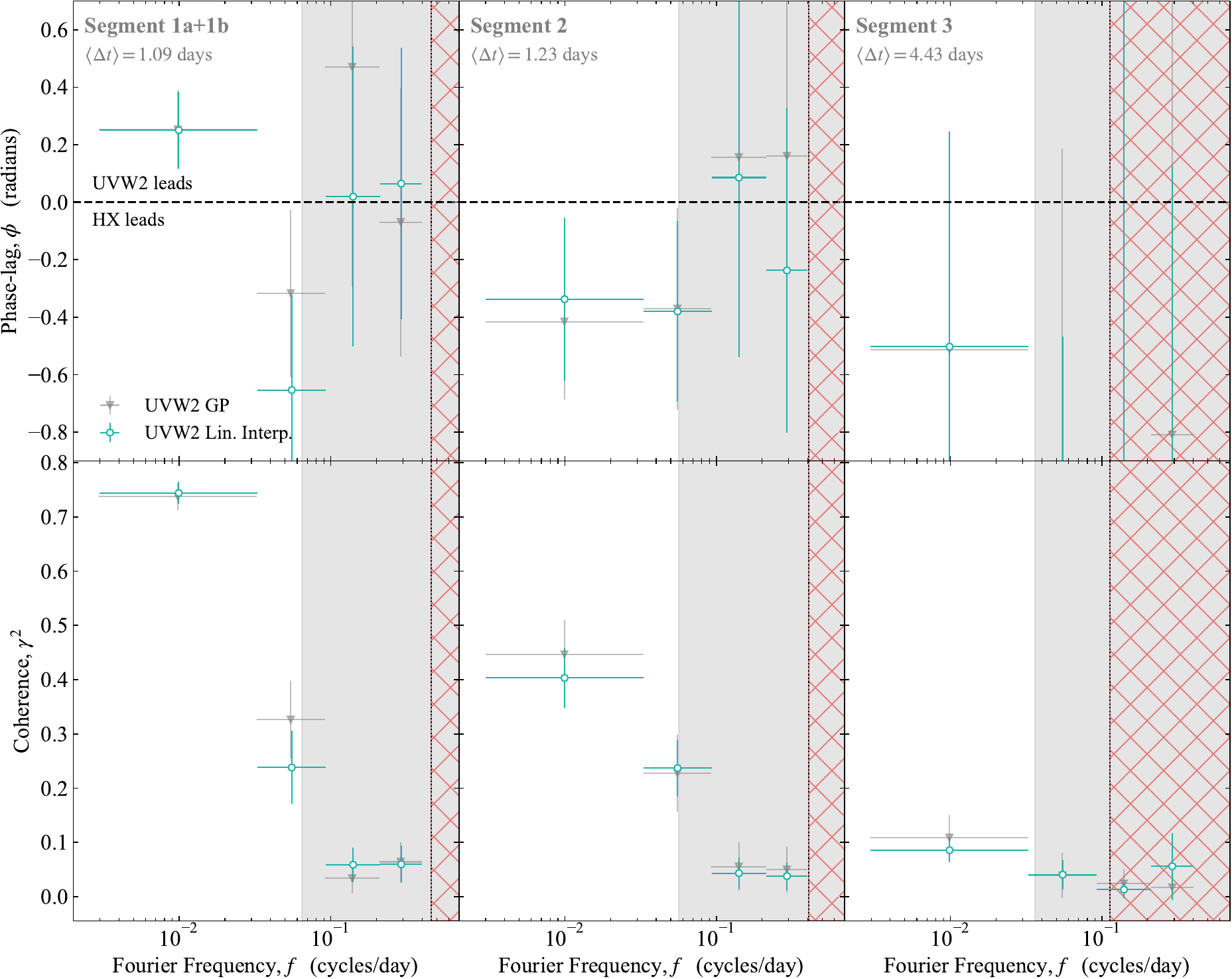}
    \caption{Comparison of the phase-lags (top) and coherence (bottom) predicted from our GPs (grey triangles) and from linear interpolation (turquoise). There is generally very little difference here.}
    \label{app_fig:LagAndCoh_linInterp}
\end{figure*}

\section{Consistency checks in the Fourier lags between GP kernels}
\label{app:GP_kerns}

\begin{figure*}
    \centering
    \includegraphics[width=\textwidth]{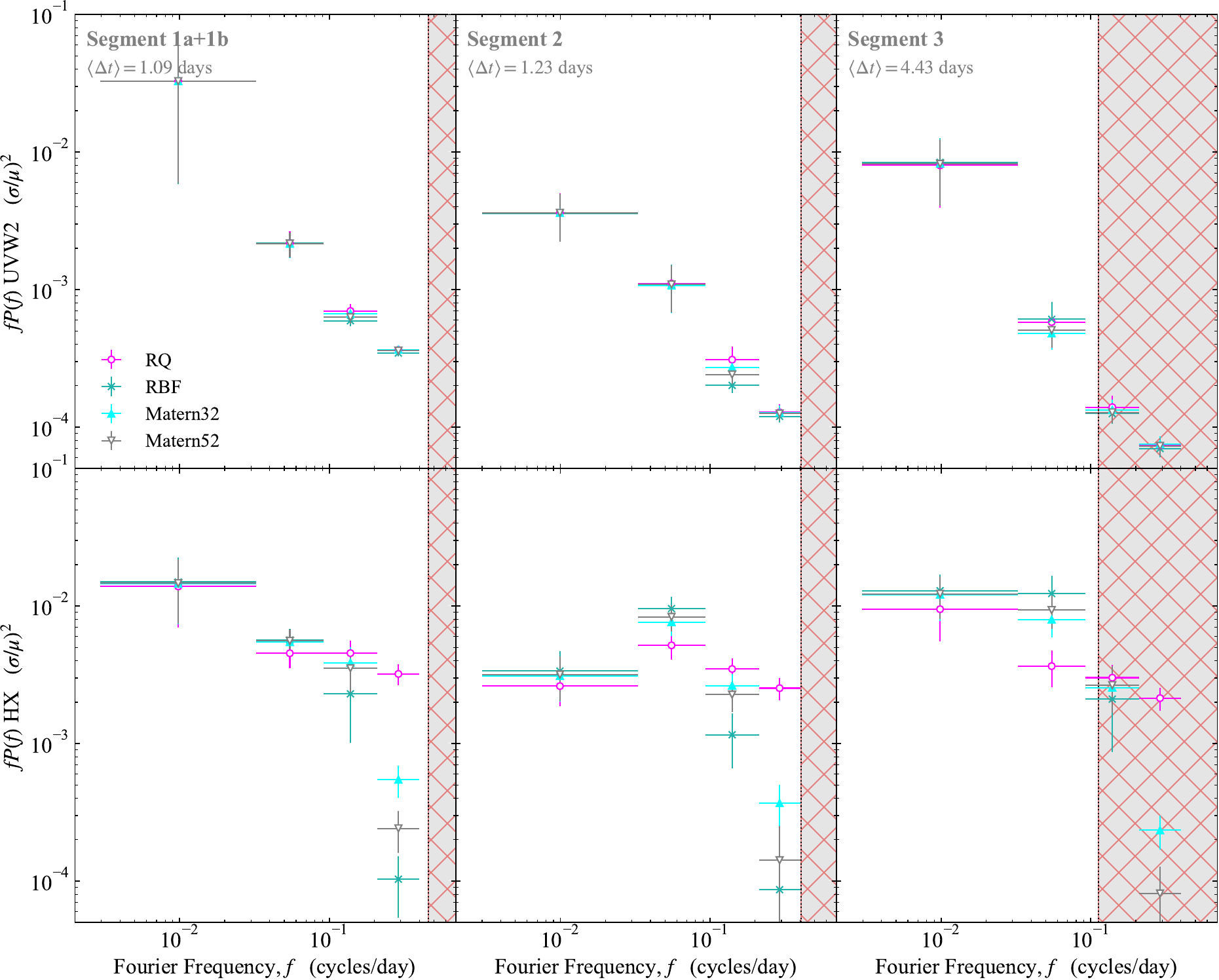}
    \caption{Comparison of the power-spectra for UVW2 (top) and HX (bottom) predicted from different GP kernels. In the UVW2 case there is typically very little difference, as this is mainly dominated by low-frequency variability. However, for the HX power-spectrum we see a clear drop in high frequency power from the RBF, Matern52, and Matern32 kernels. This is likely due to the fact that HX clearly contains a wide range of time-scales, which lead to struggles for single (or narrow) time-scale kernels.}
    \label{app_fig:power_KernTST}
\end{figure*}

\begin{figure*}
    \centering
    \includegraphics[width=\textwidth]{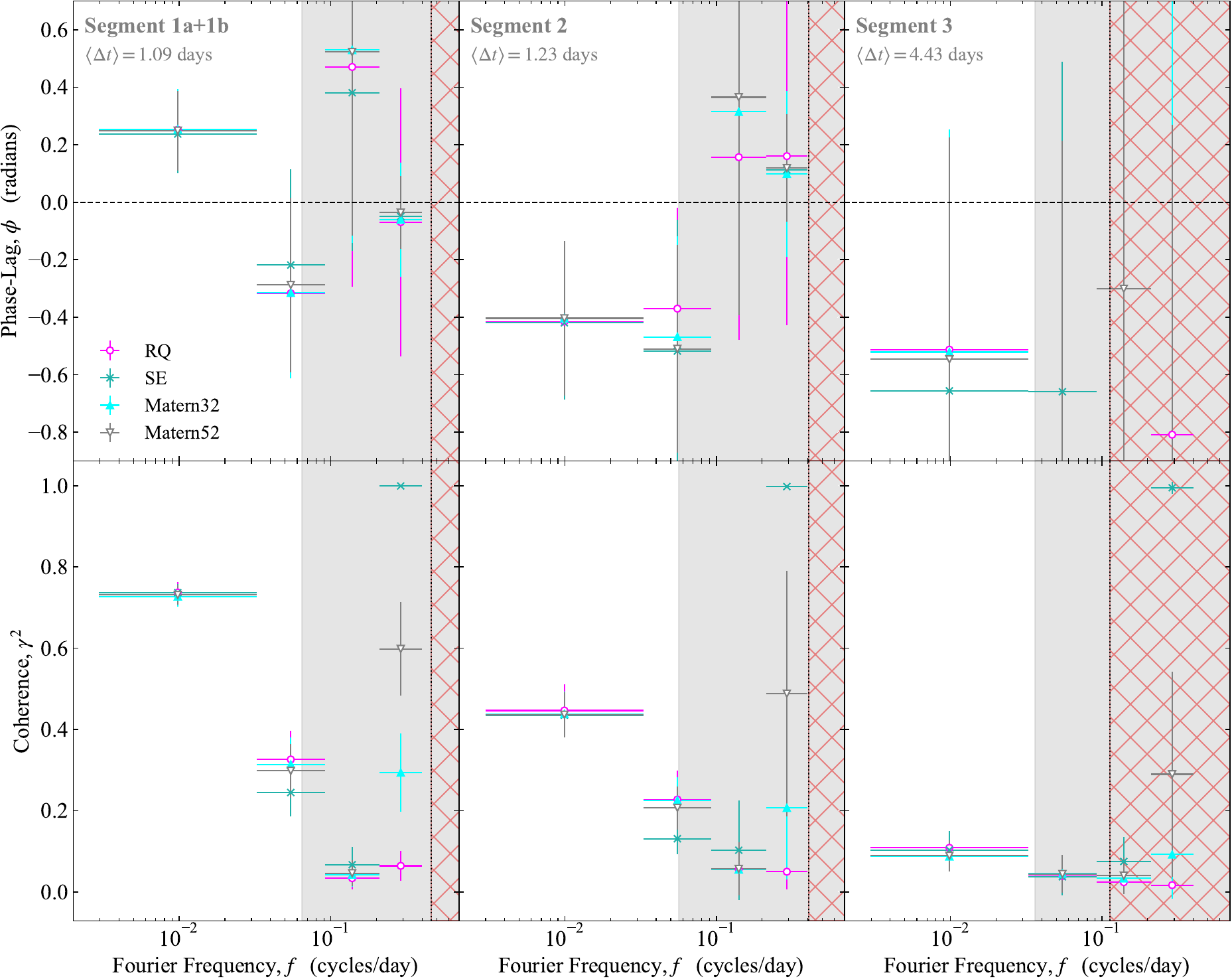}
    \caption{Comparison of the phase-lags (top) nad coherence (bottom) of different GP kernels. As with Fig.\,\ref{app_fig:power_KernTST} these are mostly consistent except at the highest frequency, where the coherence for the RBF, Matern32, and Matern52 kernels increase. This is likely due to those GPs (falsely) predicting little to no variability on these time-scales, which by its nature will be perfectly coherent.}
    \label{app_fig:LagCoh_KernTST}
\end{figure*}

{\renewcommand{\arraystretch}{1.6}
\begin{table}
    \centering
    \begin{tabular}{c c c}
        Kernel & LML (UVW2) & LML (HX) \\
        \hline
        SE & 883.2 & -27.7\\
        Matern-$\frac{3}{2}$ & 1112.6 & 162.7 \\
        Matern-$\frac{5}{2}$ & 1049.8 & 93.7 \\
        \textbf{RQ} & \textbf{1139.4} & \textbf{351.6}
    \end{tabular}
    \caption{Log marginal likelihood (LML) for the UVW2 and HX data, for different GP kernels, after optimization. Here a larger number implies a higher probability of the model given the data. For both UVW2 and HX the RQ kernel clearly wins. However, we note that HX performs significantly worse than UVW2. This is perhaps due to the wider range of time-scales present in the data (i.e flatter power-spectrum) or the significantly larger noise level.}
    \label{tab:my_label}
\end{table}
}


\bsp	
\label{lastpage}
\end{document}